\documentclass[preprint]{aastex}

\usepackage{epsfig}

\def\etal{\hbox{et al.\ }$\,$}
\def\simlt{\lower.5ex\hbox{$\; \buildrel < \over \sim \;$}}
\def\simgt{\lower.5ex\hbox{$\; \buildrel > \over \sim \;$}}
\def\kms{km s$^{-1}$\,}

\renewcommand{\sun}{\mbox{$_\odot$}}

\slugcomment{\bf Accepted by the Astrophysical Journal}

\begin{document}

\title{Optical Spectroscopy of GRO J1655$-$40} 

\author{Roberto Soria\altaffilmark{1},
   Kinwah Wu\altaffilmark{2},
   Richard W. Hunstead\altaffilmark{3}}

\altaffiltext{1}{Research School of Astronomy and Astrophysics, 
   Australian National University, Private Bag, Weston Creek Post 
   Office, ACT 2611, Australia; roberto@mso.anu.edu.au}
\altaffiltext{2}{Research Centre for Theoretical Astrophysics, 
   School of Physics, University of Sydney, NSW 2006, Australia}
\altaffiltext{3}{Department of Astrophysics, 
   School of Physics, University of Sydney, NSW 2006, Australia}

\begin{abstract}
\rightskip=\leftskip

We have obtained optical spectra of the soft X-ray transient 
GRO J1655$-$40 during different X-ray spectral states (quiescence, 
high-soft and hard outburst) between 1994 August 
and 1997 June. Characteristic features observed during the 1996--97 
high-soft state were: a) broad absorption lines at H$\alpha$ and H$\beta$, 
probably formed in the inner disk; b) double-peaked 
\ion{He}{2} $\lambda 4686$ emission lines, formed in a temperature-inversion 
layer on the disk surface, created by the soft X-ray irradiation; 
c) double-peaked H$\alpha$ emission, with a strength associated 
with the hard X-ray flux, suggesting that it was probably emitted from 
deeper layers than \ion{He}{2} $\lambda 4686$.  
The \ion{He}{2} $\lambda 4686$ line profile appeared approximately symmetric, 
as we would expect from a disk surface with an axisymmetric emissivity 
function. The Balmer emission, on the other hand, appeared to come only 
from a double-armed region on the disk, possibly the locations  
of tidal density waves or spiral shocks. The observed rotational velocities 
of all the double-peaked lines suggest that the disk was extended 
slightly beyond its tidal radius. 

Three classes of lines were identified in the spectra taken  
in 1994 August--September, during a period of low X-ray activity between 
two strong X-ray flares: broad absorption, broad (flat-topped) emission 
and narrow emission. We have found that the narrow (single-peaked 
or double-peaked) emission lines cannot be explained by a conventional 
thin accretion disk model. We propose that the system was in a transient 
state, in which the accretion disk might have had an extended optically 
thin cocoon and significant matter outflow, which would also explain the 
systematic blue-shift of the narrow emission lines and the flat-topped 
profiles of the broad emission lines.

After the onset of a hard X-ray flare the disk signatures 
disappeared, and strong single-peaked H$\alpha$ and Paschen emission 
was detected, suggesting that the cocoon became opaque to optical 
radiation. High-ionization lines disappeared or weakened.
Two weeks after the end of the flare, the cocoon appeared to be 
once again optically thin.

\end{abstract}



\section{INTRODUCTION}

Among the low-mass X-ray binaries, soft X-ray transients (SXTs) offer 
a favorable opportunity of detecting stellar-mass black-hole candidates 
(BHCs). Measurements of the binary periods and the orbital velocities 
of the companion stars in seven SXTs yield mass functions 
$\simgt 3M_{\odot}$ for their compact stars (McClintock 1998, 
Filippenko \etal 1999). 
As these masses are higher than the upper limit to the 
mass of neutron stars for known equations of state of nucleon matter, 
the compact stars in SXTs are all BHCs. 
In other two systems, the mass functions are $< 3M_{\odot}$ but the 
inclination angles are thought to be sufficiently low that the 
compact objects can also be classified as BHCs (McClintock 1998).

SXTs show transitions between different spectral states.
In a simple classification, we can identify an
``off'' (or ``quiescent'') state when the source is inactive, and a 
``high-soft state'' and a ``low-hard state'' defined by their X-ray 
spectral properties, when the source is X-ray active.
Often an increase of the optical brightness and radio flares 
accompany the X-ray outbursts. (See e.g.\ Tanaka \& Lewin 1995 
for a review of the properties of SXTs.) 

GRO~J1655$-$40 is an SXT discovered by BATSE on 1994 July 27 
(Zhang \etal 1994). Between 1994 and 1995 it underwent several X-ray 
outbursts (see Tavani \etal 1996), after which it retreated to a quiescent 
state. In late April 1996, another outburst started, and it lasted more than 
one year. The 2 -- 12 keV X-ray light curve of GRO~J1655$-$40 during the 
1996 outburst is shown in 
Figure~\ref{batse_asm} (from the quick-look results provided by the 
RXTE/ASM team); in the same figure we also show the hard (20 -- 100 keV)
X-ray flux detected by BATSE over the same period of time.

The binary period of GRO~J1655$-$40 is $P = 2.62157 \pm 0.00015$~d 
(Orosz \& Bailyn 1997). The time of 
inferior conjunction of the secondary star is 
HJD $2449838.4209 \pm 0.0055$ (Orosz \& Bailyn 1997). An alternative 
ephemeris, given by van der Hooft \etal (1998), is 
HJD $2449838.4198(52) + 2.62168(14) \times$~N.

The mass function determined from the 
kinematics of the system by Orosz \& Bailyn (1997) implies a mass 
$M_{1} = (7.0 \pm 0.2)~M_{\sun}$ for the compact object. 
By taking into account the effect of X-ray irradiation of the secondary 
star, Phillips, Shahbaz \& Podsiadlowski (1999) derived
lower values for the mass of the primary: they obtained  
$4.1 < M_{1} < 6.6 M_{\sun}$ (90 per cent confidence level). 
By using only data taken during an X-ray quiescent state, Shahbaz \etal 
(1999) obtained $5.5 < M_{1} < 7.9 M_{\sun}$ (95 per cent confidence level). 
In any case, the mass of the compact object makes GRO~J1655$-$40
a candidate black hole binary.  The major 
orbital parameters of the system are listed in Table 1.  

During the quiescent states the system is optically faint, with 
$V \approx 17$ mag (Bailyn \etal 1995b), but it can 
brighten substantially when an X-ray outburst occurs. In August 1994, it  
reached $V = 14.4$ (Bailyn \etal 1995a), 
and in May 1996, $V = 15.4$ (Horne \etal 1996).

Since its discovery, we have observed GRO~J1655$-$40 spectroscopically 
on four occasions: 1994 August -- September, 1996 April, 
1996 June and 1997 June. The observations covered 
periods in which the system was in a pre-outburst quiescence, at the 
onset of a hard X-ray flare and in a high-soft state. In a previous 
paper (Soria \etal 1998) we presented the orbital-phase-dependent 
velocity shifts of the \ion{He}{2} $\lambda 4686$ and 
\ion{N}{3} $\lambda \lambda 4641, 4642$ emission lines. 

In this paper we extend our study to include the Balmer lines and the orbital 
variations in their line profiles.   
Technical details of our observations are outlined in \S2. In \S3  
we present the results of the spectroscopic observations we conducted 
in 1996 April, and in \S4 we present those we obtained in 1996 June 
and 1997 June. In \S5 we discuss the profiles of the Balmer lines
observed in the 1996 -- 1997 soft X-ray outburst, and we attempt to 
locate the emission regions. We also discuss some 
physical mechanisms responsible for line emission and absorption in the disk.
In \S6 we present the results of the observations we carried out during 
the 1994 hard X-ray outburst. We outline the general spectral features, 
we discuss the origin, intensity and profile of the main absorption 
and emission lines, and we compare the two sets of spectra from 1994 and 1996.
In particular, we show the evolution of the optical spectrum before, 
during and after a major hard X-ray flare.

\section{OBSERVATIONS}

GRO~J1655$-$40 was observed by one of us (RWH) as a target of 
opportunity on each night from 1994 August 30 to 1994 September 4, 
with the RGO spectrograph and Tektronix 1k $\times$ 1k thinned CCD 
on the 3.9~m Anglo-Australian Telescope (AAT). 
Spectra were obtained in two regions, $6278-6825$ \AA, 
centered on the H$\alpha$ line, and $4432-5051$ \AA, 
covering the \ion{N}{3}, \ion{He}{2} and H$\beta$ lines. 
Gratings with 1200 grooves/mm were used, with the blaze direction 
oriented towards the 25 cm camera, giving a resolution of 
1.3 \AA\ FWHM. On September 6, we obtained simultaneously a blue 
($3925-5500$ \AA) spectrum with the RGO spectrograph 
and 600 grooves/mm gratings, at a resolution of 2.5 \AA\ FWHM, 
and a red ($5500-11000$ \AA) spectrum 
with the FORS spectrograph, via a dichroic beam splitter, 
at a resolution of 20 \AA\ FWHM. The latter observations 
were kindly obtained for us by Paul Francis.

In 1996 April 20 -- 21 we carried out a scheduled observation with 
the Double Beam Spectrograph (DBS) on the ANU 2.3~m Telescope at Siding 
Spring Observatory. The detectors on the two arms of the spectrograph 
were SITe 1752$\times$532 CCDs. Gratings with 300 grooves/mm were used 
for the blue ($3600-5700$ \AA) and the red ($5700-9300$ \AA) bands, 
and low-resolution spectra (resolution $=4.8$ \AA\ FWHM) were 
obtained. More extensive observations were carried out 
in 1996 June 8 -- 12 and June 17, and 1997 June 14 -- 15, 
with the DBS on the ANU 2.3~m Telescope. 
1200 grooves/mm gratings were used for both the blue 
($4150-5115$ \AA) and the red ($6300-7250$ \AA) bands, giving a
resolution of 1.3 \AA\ FWHM.    
We list our observations in Table 2.

\section{PRE-OUTBURST STATE (1996 APRIL)} 

\subsection{Overview} 

GRO~J1655$-$40 was in quiescence between late 1995 August and 1996 April. 
In 1996 March, its soft X-ray luminosity, inferred from the 
ASCA observations (at $2-10$ keV), was $\approx 2\times 10^{32}$ erg s$^{-1}$ 
(see Orosz \& Bailyn 1997). The source was not detected by RXTE/ASM 
(at $2-12$ keV) above the intensity level of $\approx 12$ mCrab
before 1996 April 25 (Levine \etal 1996; 
Remillard \etal 1996); it was also undetected by BATSE 
before May. On April $25.38 \pm 0.78$ UT (HJD $2450198.88 \pm 0.78$), the 
soft ($2-12$ keV) X-ray intensity began to rise (Remillard \etal 1996). 
The optical brightening, however, had already started on 
April 20 (Orosz \etal 1997). The fitted times of the initial rise 
were April $19.25 \pm 0.29$ UT (HJD $2450192.76 \pm 0.29$) for the 
$I$ band, April $19.37 \pm 0.26$ UT (HJD $2450192.88 \pm 0.26$) for the 
$R$ band, April $19.82 \pm 0.15$ UT (HJD $2450193.32 \pm 0.15$) for the 
$V$ band and April $20.34 \pm 0.18$ UT (HJD $2450193.84 \pm 0.18$) for 
the $B$ band. 

Our spectroscopic observations were conducted between HJD $2450194.26$ 
and HJD $2450194.30$ (April 20), and between 
HJD $2450195.10$ and HJD $2450195.15$ (April 21),  
just after the initial rise in the optical brightness, but before the 
RXTE/ASM detected the rise in the X-ray intensity. 
On the first night we took a series of five 
600-s spectra and one 1200-s spectrum (although focussing problems on the red 
arm of the spectrograph affected the quality our red spectra).
The binary phase during the 
observations was $0.485 < \phi < 0.501$ (ephemeris of 
Orosz \& Bailyn 1997), and the inferior conjunction of the secondary 
star was at $\phi = 0.75$. On the following night, we took seven 
600-s spectra, at the binary phase $0.804 < \phi < 0.824$.  
(If we adopt the ephemeris of van der 
Hooft \etal (1998), the observations were made at phase 
$0.480 < \phi < 0.496$ on the first night, and 
$0.799 < \phi < 0.819$ on the second.)

\subsection{General spectral features} 

The \ion{H}{1} Balmer and Paschen lines are clearly seen 
in absorption in our spectra (Figures~\ref{fluxspectra} and \ref{96apr}).   
The \ion{He}{2} $\lambda 4686$ emission line,  
the Bowen fluorescence \ion{N}{3} $\lambda 4634$ line and the 
blend \ion{N}{3} $\lambda \lambda 4641, 4642$ were not detected on either 
night. These emission lines, which are often seen from X-ray 
irradiated accretion disks (Smak 1981; Horne \& Marsh 1986), were, 
however, observed 20 days later (on May 11) by Hynes \etal (1998b). 

By comparing our spectra with the 
spectra of F giant and subgiant stars and with the spectra of the system 
taken by Orosz \& Bailyn (1997) during a quiescent state in February 1996, 
we conclude that the optical continuum emission was predominantly 
from the secondary star. 
This conclusion is supported by the agreement between the observed 
velocity shifts of the Balmer lines in our spectra and the 
radial velocity of the secondary star expected at those phases 
(not shown).

\section{HIGH-SOFT STATE (1996--1997)} 

\subsection{1996 June observations} 
 
\begin{figure}  
\begin{tabular}{c}
\epsfxsize=65mm\epsfbox{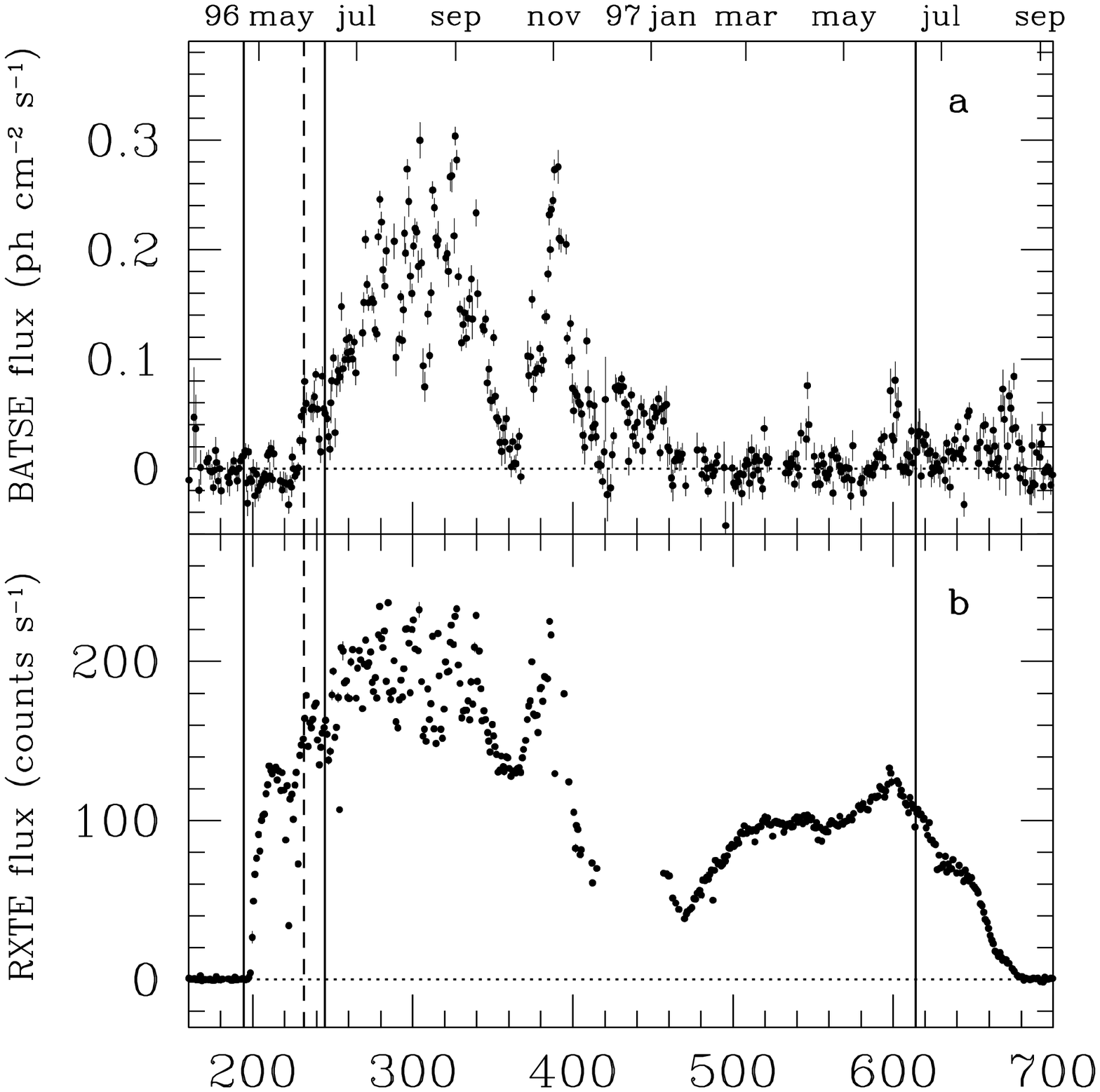}\\
\epsfxsize=65mm\epsfbox{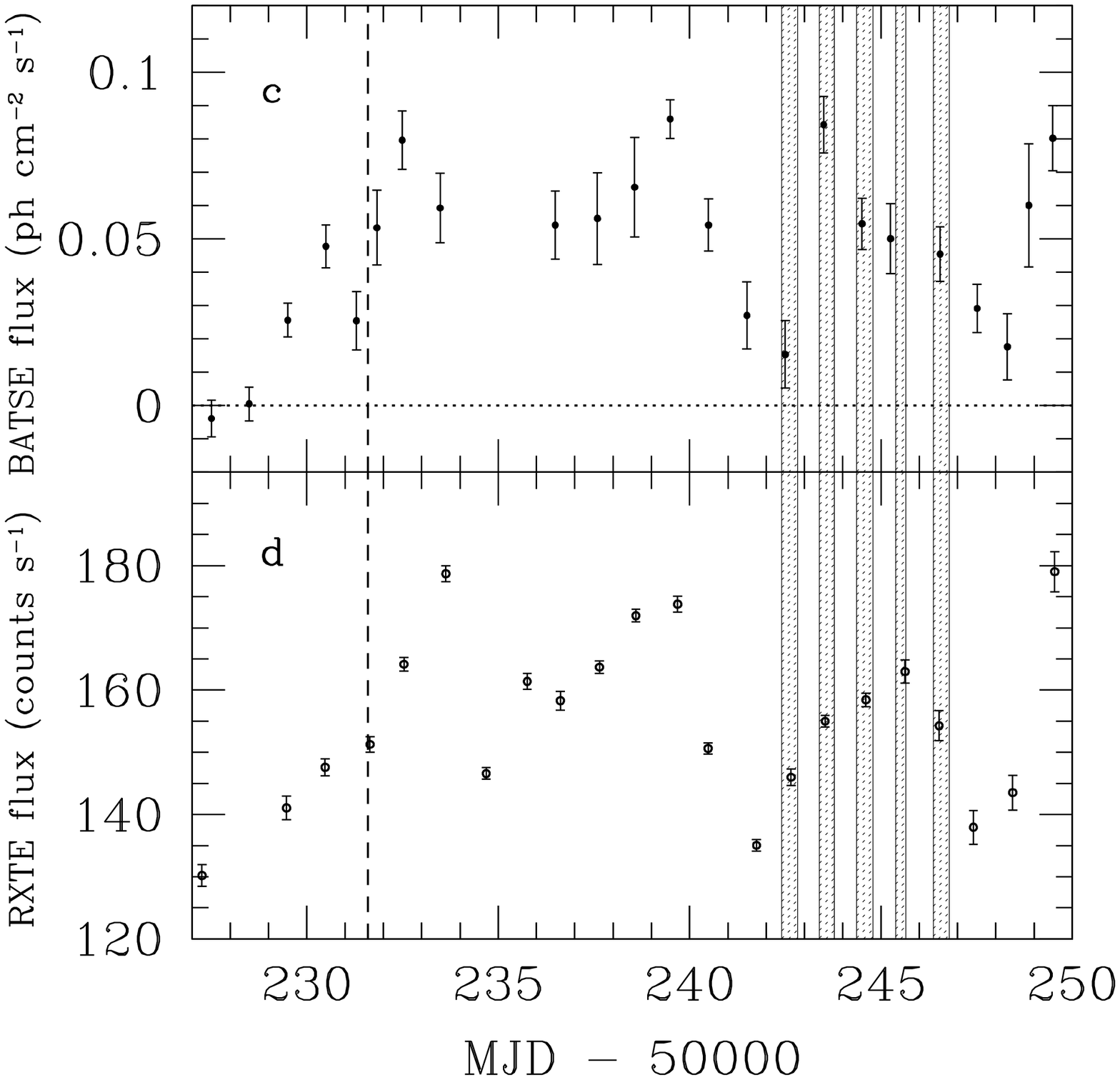}\\  
\end{tabular}
\caption{a: daily averages of the hard ($20-100$ keV) X-ray flux 
measured by BATSE. b: daily averages of the soft ($2-12$ keV) 
X-ray flux measured by RXTE/ASM.
The lightcurves cover the period from 1996 March 18 to 
1997 September 9. The solid lines mark the epochs of our 
observations (1996 April 20--21, 1996 June 8--12, 1997 June 14--15).
A transition from quiescence to a high-soft X-ray spectral state 
occurred around MJD 50198 (April 25). The X-ray spectrum hardened 
after MJD 50230 (May 27). A radio flare was observed by MOST on 
MJD 50232 (May 28.60 UT) (dashed line). Hjellming (1997) 
estimates that the peak of the radio flare occurred sometime between MJD 
50222.5 and MJD 50230.5.
The actual onset of the radio 
flare is not well determined: its flux density was already declining 
with an e-folding time of 1.4 d
when the flare was detected (Hunstead \etal 1997). 
In the lower panels, we show an expanded view of the hard (c) and 
soft (d) X-ray flux around the time of our 1996 June observations 
(shaded regions).}
\label{batse_asm}
\end{figure}

\begin{figure}
\epsfxsize=145mm\epsfbox{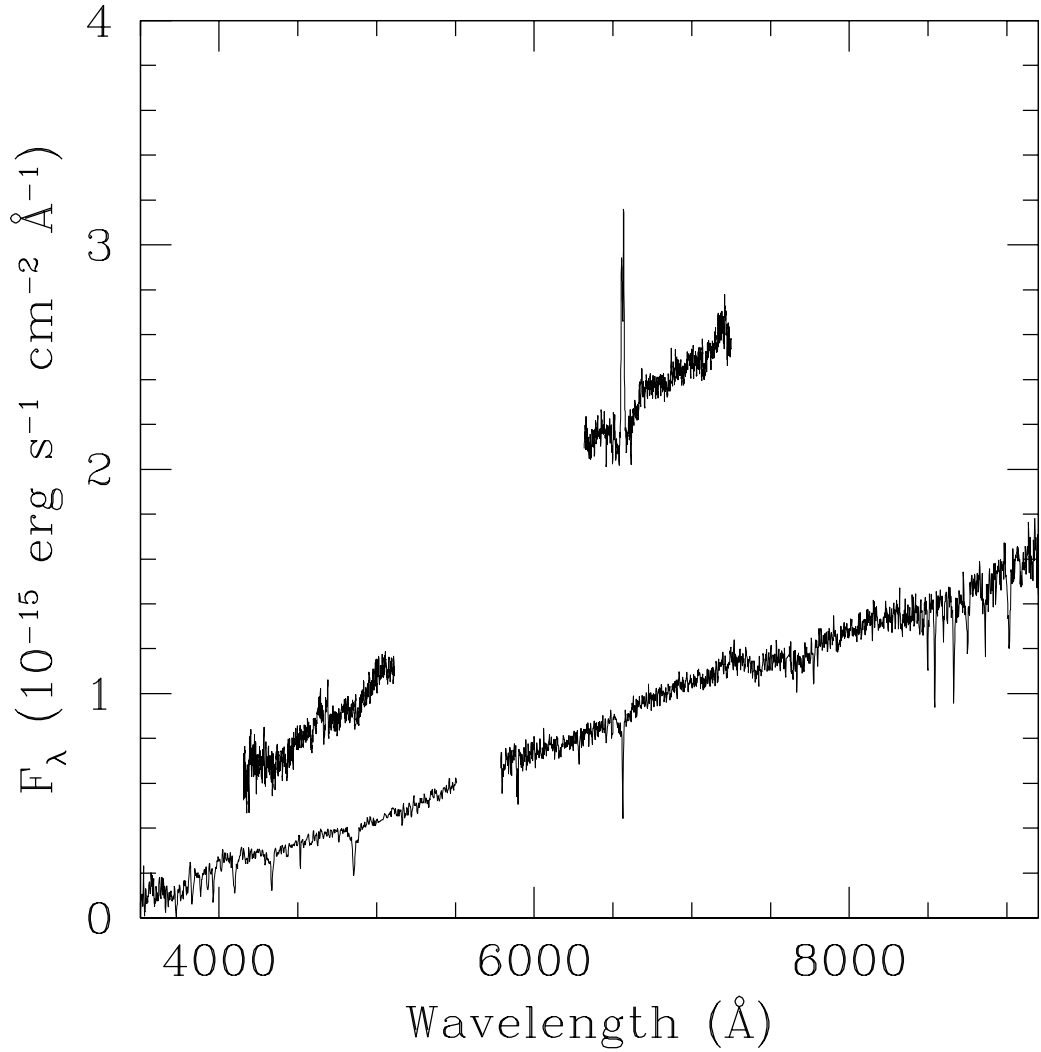}
\caption{Flux-calibrated spectra taken on 1996 April 21 (lower spectrum)
and on 1996 June 10 (upper spectrum) 
with the ANU 2.3~m telescope at Siding Spring Observatory. 
Wavelengths are vacuum heliocentric (cf.\ Figure 2 in Hynes \etal 1998b).}
\label{fluxspectra}
\end{figure}

\begin{figure} 
\begin{tabular}{c}
\epsfxsize=95mm\epsfbox{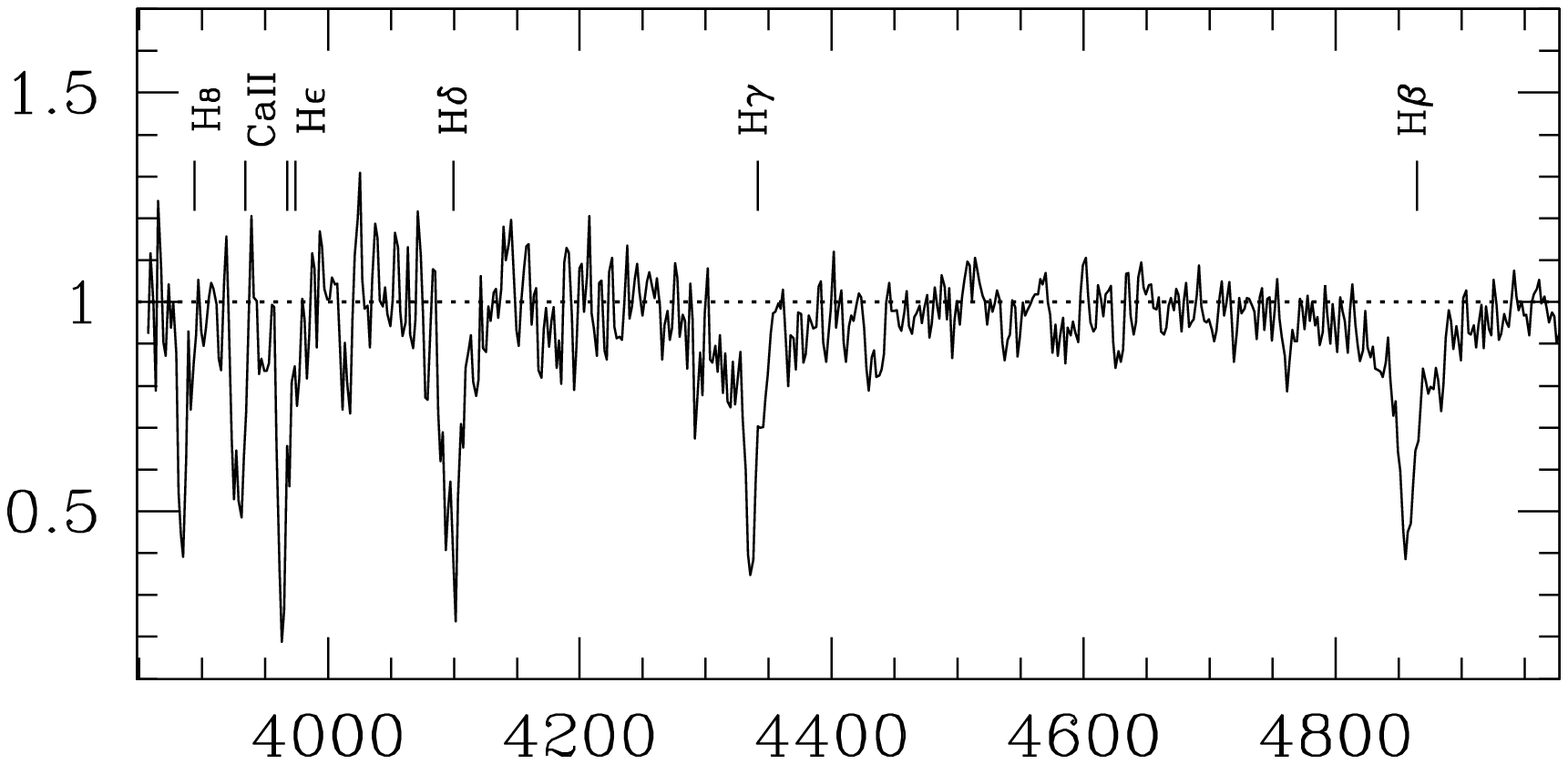}\\
\epsfxsize=95mm\epsfbox{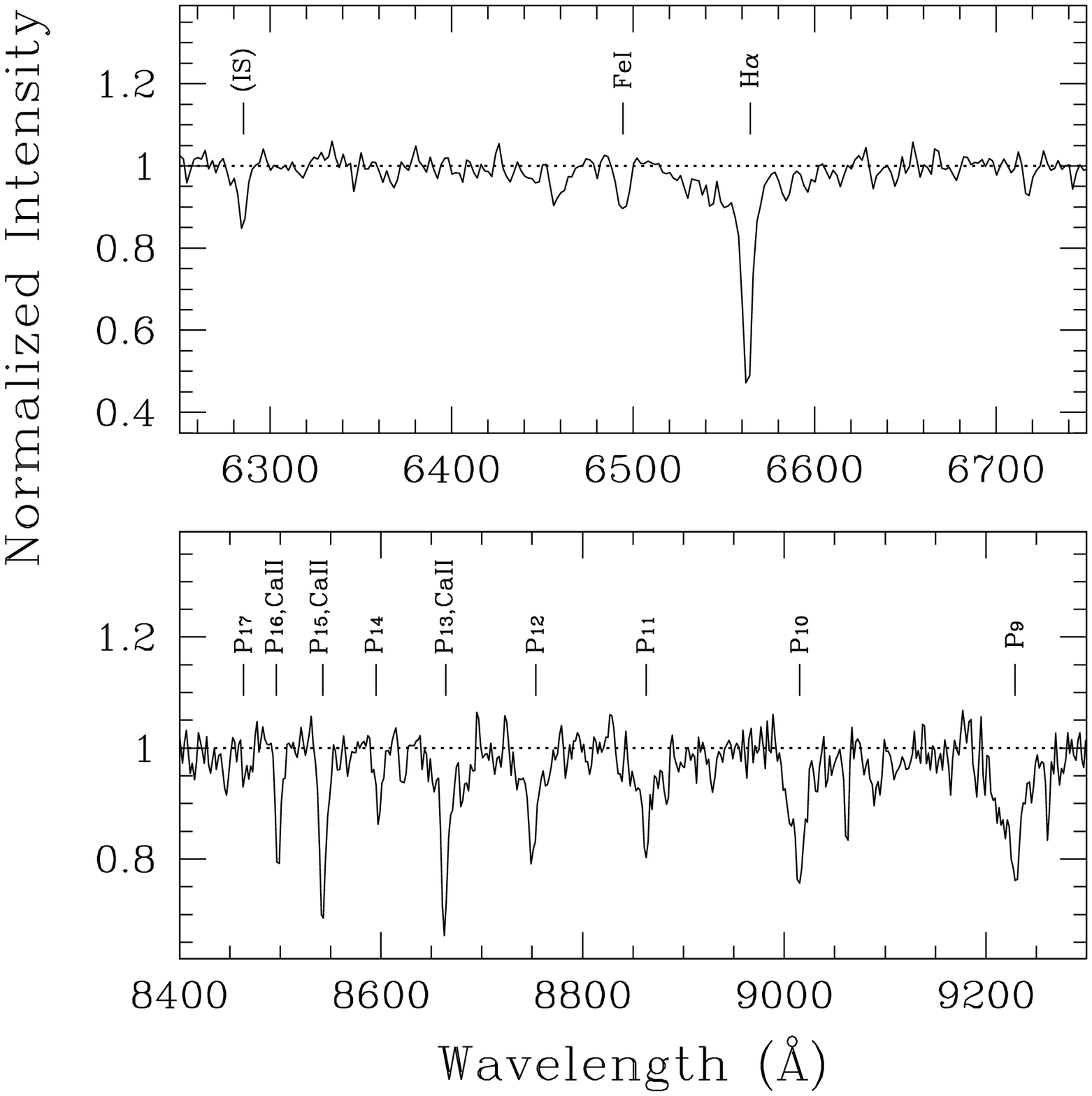}\\
\end{tabular}
\caption{Top panel: averaged blue spectrum  
obtained on 1996 April 20, at spectroscopic 
phases $0.485 < \phi < 0.501$, according to the ephemeris 
of Orosz \& Bailyn (1997). Central and bottom panels: portions 
of the averaged red spectrum 
obtained on 1996 April 21, at spectroscopic phases $0.804 < \phi < 0.824$.  
Wavelengths are vacuum heliocentric, and the 
intensities are normalized to the continuum. The Balmer series is 
in absorption. Other strong absorption features are 
\protect\ion{H}{1} Paschen series, 
\protect\ion{Ca}{2} $\lambda \lambda 3934$,$3968$ (the latter 
is blended with H$\epsilon$ $\lambda 3970$), and the triplet 
\protect\ion{Ca}{2} $\lambda \lambda \lambda 8542,8662,8498$ 
(blended with the Paschen lines).  
No emission is observed from the \protect\ion{N}{3}/\protect\ion{C}{3} lines, 
nor from \protect\ion{He}{2} $\lambda 4686$.}
\label{96apr}
\end{figure}

In late 1996 April, a transition from quiescence  
to a high/soft X-ray state occurred. The soft ($2-12$ keV) X-ray flux 
(measured by RXTE/ASM) increased rapidly, reaching $\simgt 1.5$~Crab 
and remaining above that level for more than six months.   
Around 1996 November the soft X-ray intensity began to decline; but in 
1997 January it increased again (Figure~\ref{batse_asm}). It then 
stayed at the $\approx 1.3$~Crab level till 1997 July, after which the 
system entered another quiescent state.

At the initial stage of the outburst the X-ray spectrum consisted 
of a blackbody component and a truncated power-law tail. The black body 
temperature was $T \simeq 0.8$ keV. The photon 
index of the power-law tail was $\Gamma \simeq 3.5$ and the exponential 
cutoff was at $\simeq 13$ keV (Hynes \etal 1998b). Around 
1996 May 27 -- 28 (HJD 2450230 -- 31) the X-ray spectrum hardened, 
and the source became detectable by BATSE (Figure~\ref{batse_asm}).  
The photon index of the power-law was $\Gamma \simeq 2.5$ on June 20,  
and there was no obvious sign of a high-energy cutoff below 
$\simeq 100$~keV (Hynes \etal 1998b).  
A radio flare was observed by the Molongo Observatory Synthesis 
Telescope (MOST), apparently coincident the X-ray spectral transition 
(Hunstead \& Campbell-Wilson 1996; Hunstead, Wu \& Campbell-Wilson 1997).

The optical continuum had increased a few days before the onset of the 
soft X-ray outburst (see \S3.1). After reaching a peak in mid-May, the 
optical brightness then declined steadily, with an e-folding time of 
about 50 days (Hynes \etal 1998b). By August 1996 it had returned to 
the previous quiescent level. Flickerings on time-scales of a few 
seconds, apparently correlated with X-ray variability, were reported by 
Hynes \etal (1998a). 

Our observations were carried out on June 8 -- 12 and June 17,  
during the soft X-ray outburst (Figure~\ref{batse_asm}). We  
obtained fifty simultaneous red and blue spectra,  
each of 2000 s duration. In Figure~\ref{96jun} we show   
two blue and red spectra taken on June 8 and June 10.    
The intensity of the optical continuum in May was  
slightly higher than in June (cf. Hynes \etal 1998b).  
The soft X-ray flux was similar in these two epochs; however, 
the hard X-rays turned on after May 27.

\subsection{General spectral features in the 1996 June observations}  

\begin{figure} 
\begin{tabular}{c}
\epsfxsize=65mm\epsfbox{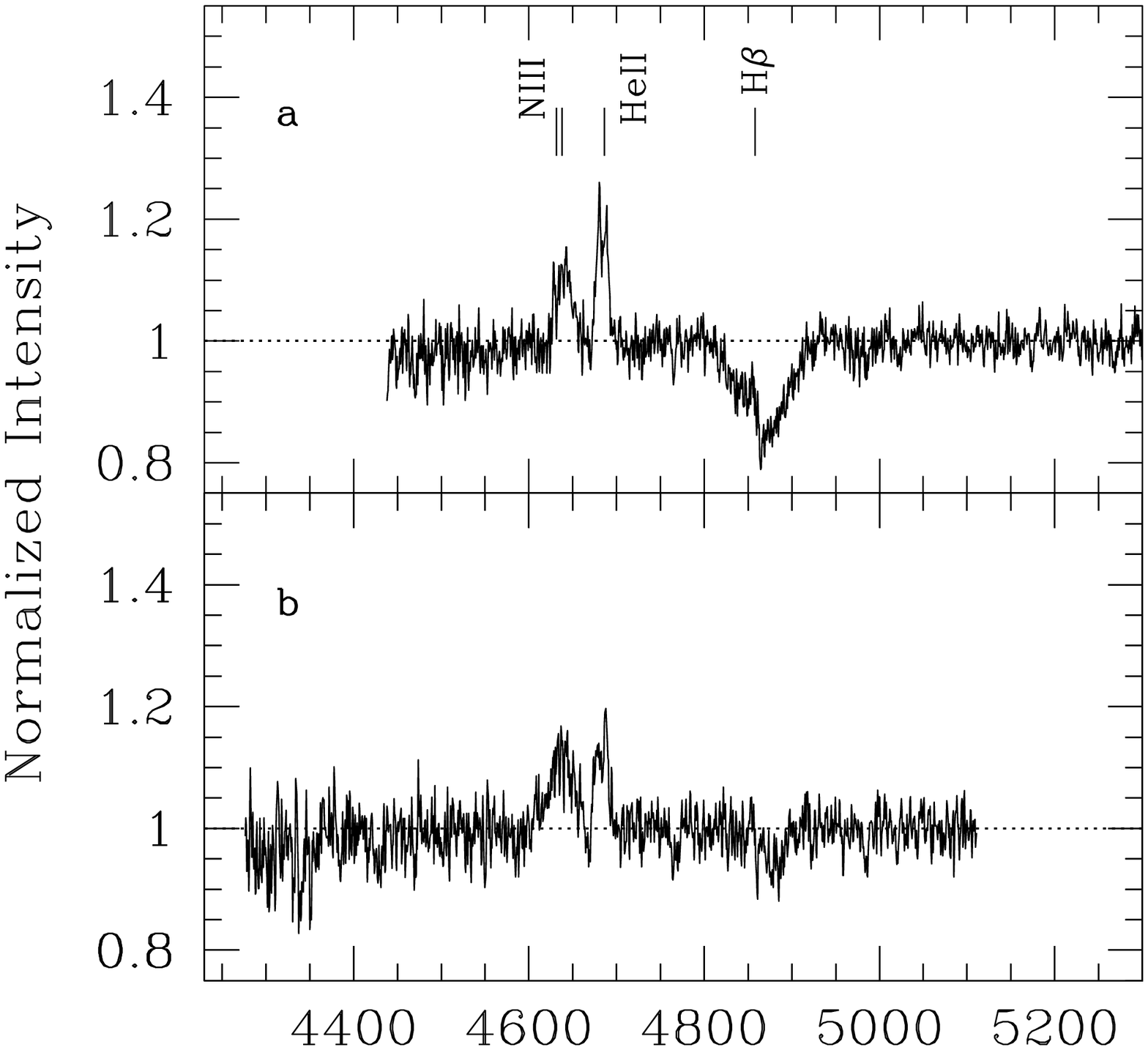}\\
\epsfxsize=65mm\epsfbox{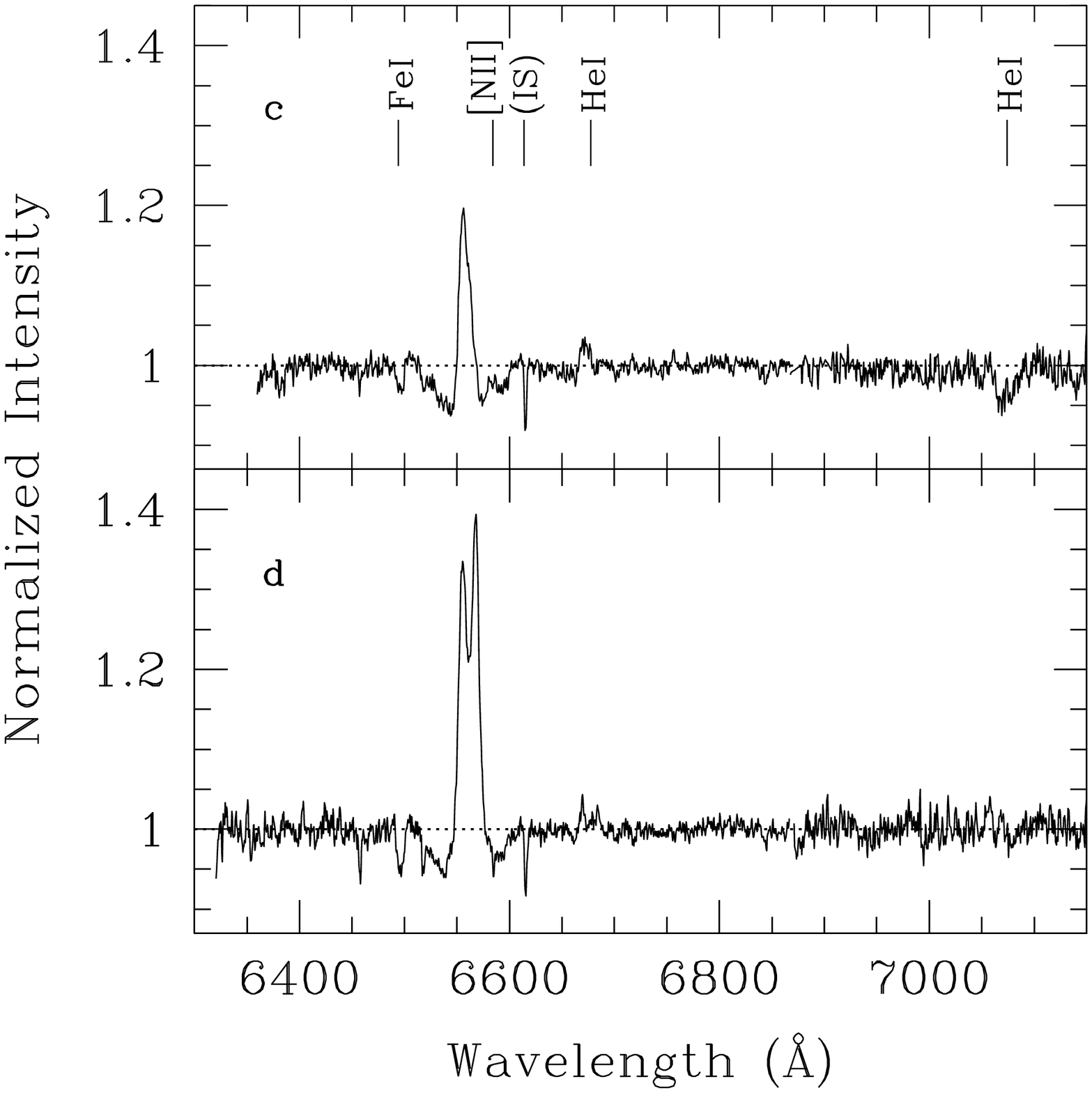}\\
\end{tabular}
\caption{a: normalized blue spectrum taken on 1996 June 8, around  
spectroscopic phase $\phi = 0.10$. b: normalized blue spectrum 
taken on 1996 June 10, around $\phi = 0.85$.
c: normalized red spectrum taken on 1996 June 8, around  
$\phi = 0.10$. d: normalized red spectrum taken 
on 1996 June 10, around $\phi = 0.85$. 
Wavelengths are vacuum heliocentric. 
Broad absorption troughs were detected at H$\beta$ and H$\alpha$ on June 8;
they were partly filled on June 10. A broad absorption 
feature at \protect\ion{He}{1} $\lambda 7065$ was also present in the June 8 
spectrum, but had disappeared on June 10. Emission was observed from 
the Bowen fluorescence lines of \protect\ion{N}{3} (single-peaked) and 
from \protect\ion{He}{2} $\lambda 4686$, H$\alpha$ and \protect\ion{He}{1} $\lambda 6678$ 
(double-peaked).
The H$\alpha$ emission was much stronger on June 10 than on June 8.
The peak-to-peak separations for \protect\ion{He}{2} $\lambda 4686$ lines 
were $525 \pm 25$ \kms on June 8, and $545 \pm 25$ \kms on June 10.
The peak-to-peak separation for H$\alpha$ was $550 \pm 10$ \kms on June 10; 
on the same night, the peak-to-peak separation for \protect\ion{He}{1} $\lambda 6678$ 
was $650 \pm 50$ \kms.
}
\label{96jun}
\end{figure}

The H$\alpha$ and H$\beta$ lines showed a broad absorption trough in 
1996 June. The absorption component was stronger on 
June 8 (EW $\simeq -6$ \AA, 
FWHM $\simeq 70$ \AA\ $\simeq 3000$ \kms for H$\alpha$; 
EW $\simeq -10$ \AA, 
FWHM $\simeq 70$ \AA\ $\simeq 4000$ \kms for H$\beta$)  
than on the other nights.    
The absorption troughs were partly filled by narrower 
emission components (Figure~\ref{96jun}), which 
were not seen in the spectra obtained by Hynes \etal (1998b) four weeks 
earlier.   

\begin{figure}  
\epsfxsize=145mm\epsfbox{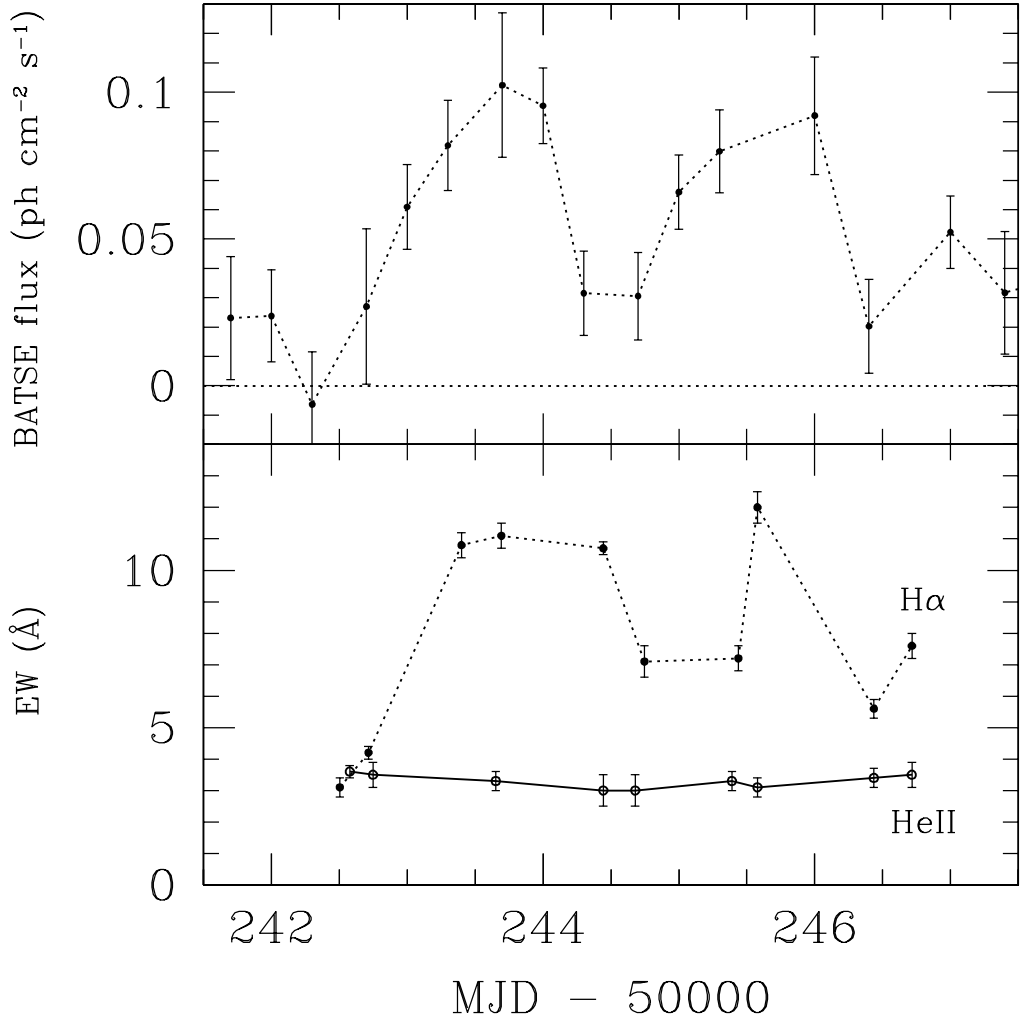}  
\caption{Top panel: hard ($20-100$ keV) X-ray flux measured by BATSE 
during the time of our 1996 June observations. Bottom panel: 
equivalent width of the H$\alpha$ and \protect\ion{He}{2} $\lambda 4686$ 
emission lines. No correlation was found beween the hard X-ray flux and 
the EW of \protect\ion{He}{2} $\lambda 4686$, but there is evidence of a 
correlation between the hard X-ray flux and the H$\alpha$ emission. 
A similar correlation was observed in 1994 August -- September 
(cf.\ Figure~\ref{batse_EW94}). 
}
\label{batse_EW1}
\end{figure}

\begin{figure}  
\epsfxsize=145mm\epsfbox{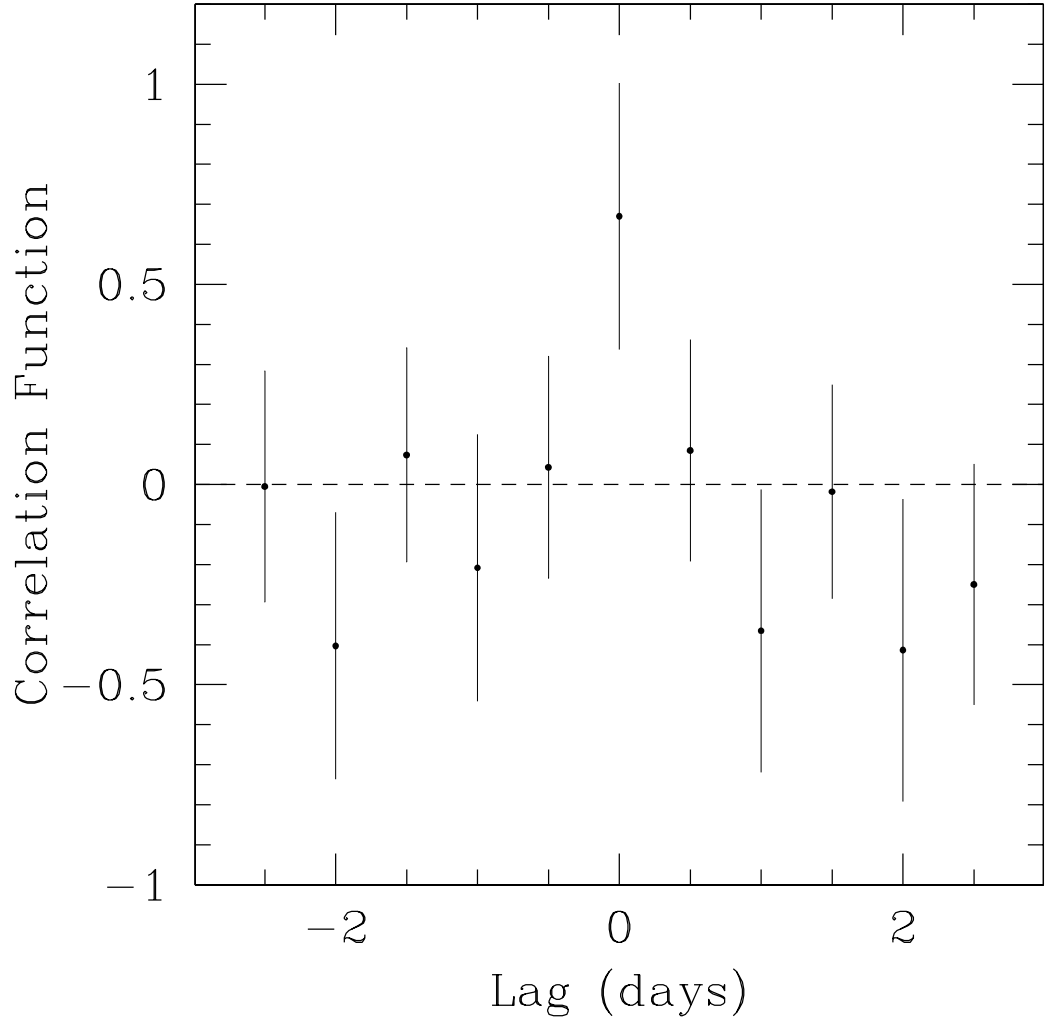} 
\caption{Discrete cross-correlation function for the equivalent width 
of the H$\alpha$ emission line and the hard ($20-100$ keV) X-ray 
flux in 1996 June. The error is large, mainly because of the limited 
number of datapoints available for the correlation function; nonetheless, 
the two quantities appear correlated. Better evidence of a correlation 
between hard X-ray flux and Balmer emission was found in our 1994 
observations (cf.\ Figure~\protect\ref{batse_EW94}).
}
\label{lag}
\end{figure}

The equivalent width of the H$\alpha$ emission seemed to 
increase with the hard X-ray flux (Figure~\ref{batse_EW1}).
The emergence of the H$\alpha$ emission component after the 
May 27 turn-on of the hard X-ray flux also suggests a direct association 
between these emissions. We have calculated the discrete correlation 
function, defined by Edelson \& Krolik (1988), for 
the EW of the H$\alpha$ emission line and the hard X-ray flux and  
have found a 2-$\sigma$ peak in the correlation function at the lag 
of $0.0 \pm 0.5$ d, for a 0.5 d interval bin (Figure~\ref{lag}). 
A similar correlation is also found in our 1994 August -- September  
data (see \S6.4.1).   

The average H$\alpha$ emission profile was double-peaked, with 
FWHM $\simeq 900$ \kms, and peak-to-peak velocity separation 
$= 500 \pm 50$ \kms. However, the relative strength of the two peaks 
varied substantially over each orbital cycle (see \S4.3).

\begin{figure}
\epsfxsize=145mm\epsfbox{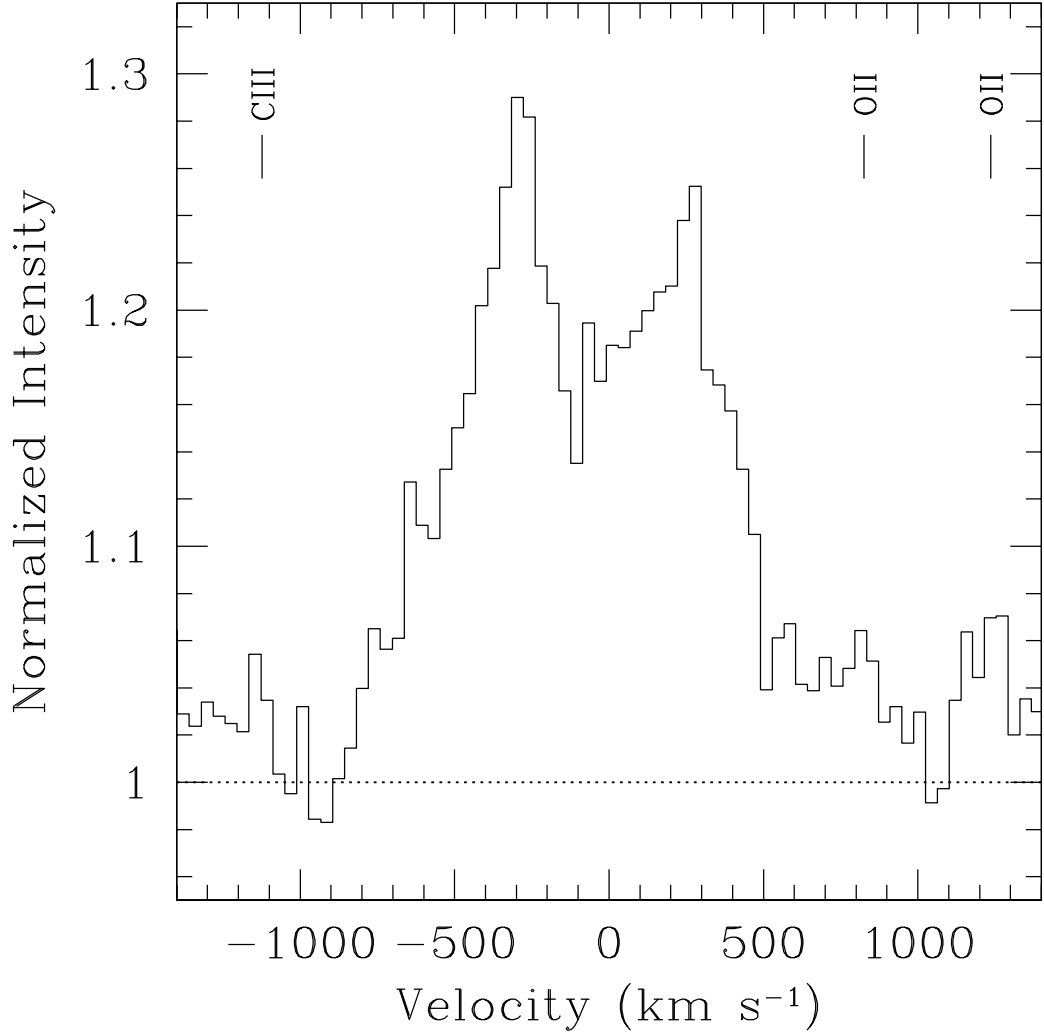} 
\caption{Normalized line profile of \protect\ion{He}{2} $\lambda 4686$, 
obtained by averaging the spectra taken on 1996 June 8 at orbital phases 
$0.08 < \phi < 0.12$. Wavelengths are vacuum heliocentric 
and the intensity is normalized to the continuum. The velocity 
zeropoint is the systemic velocity 
($\gamma = -142.4 \pm 1.6$ \kms, see Table 1). Weak emission is also detected 
at positions consistent with \protect\ion{C}{3} $\lambda 4664$, \protect\ion{O}{2} 
$\lambda 4699$ and \protect\ion{O}{2} $\lambda 4705$, as marked in the figure.}
\label{HeII_profile96}
\end{figure}

The \ion{He}{2} $\lambda 4686$ emission line had a double-peaked 
profile (Figure~\ref{HeII_profile96}) with a peak-to-peak 
separation of $8.5 \pm 0.5$ \AA, corresponding to a velocity separation of 
$545 \pm 30$ \kms. A double-peaked \ion{He}{2} $\lambda 4686$ emission 
line had been detected before the X-ray hardening (see Table 6 
in Hynes \etal 1998b): archival spectra from the Anglo-Australian 
Telescope show that the peak-to-peak velocity separation 
was $480 \pm 40$ \kms on 1996 May 11. The central position of the line varied 
sinusoidally, with the variations consistent with the projected 
radial velocity of the primary star for a mass ratio $q = 0.33$  
(Soria \etal 1998), suggesting that the emission originated 
from the accretion disk. The average line center was blue-shifted 
by $40 \pm 10 \pm 5$ \kms with respect to the systemic velocity 
determined by Orosz \& Bailyn (1997), where 
the first source of error is in the determination of the line center, 
the second is the systematic error in the wavelength calibration.
We have not found any significant correlation between 
the EW of the \ion{He}{2} $\lambda 4686$ line and the hard X-ray 
flux (Figure~\ref{batse_EW1}).    
 
The \ion{He}{1} $\lambda 6678$ emission line appeared to be double-peaked, 
with only one peak visible at some orbital phases. The velocity separation, 
which was determined when the two peaks were present, varied between $520$ 
and $650$ \kms over the time of our observations. These velocities are 
consistent with those of the H$\alpha$ and the \ion{He}{2} $\lambda 4686$ 
lines. The EW of the line was about $0.7$ \AA. 

A weak narrow absorption component was also detected at $\lambda 6678$ 
(not shown), superimposed on the broader, double-peaked emission component. 
The narrow absorption component was visible at most 
orbital phases and had an EW $\simeq -0.2$ \AA\ and an 
FWHM $\simeq 4$ \AA\ ($\simeq 170$ \kms). A similar narrow absorption 
component was also found at $\lambda 7065$. 

\begin{figure}   
\epsfxsize=145mm\epsfbox{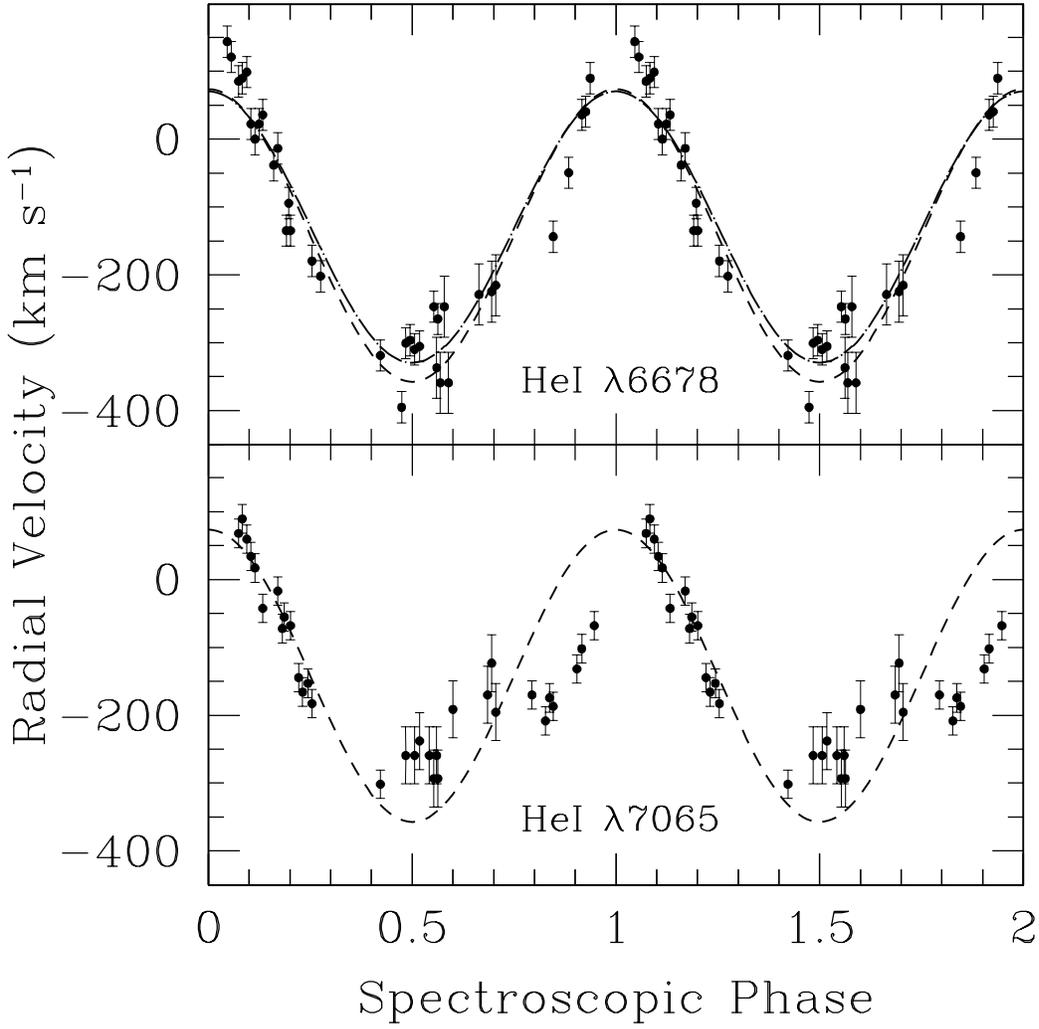}  
\caption{Radial velocity shifts of the narrow absorption components  
observed at \protect\ion{He}{1} $\lambda 6678$ (top panel) and \protect\ion{He}{1} 
$\lambda 7065$ (bottom panel) in 1996 June. 
Dash-dotted line (top panel): best sinusoidal fit to the velocity shifts of 
\protect\ion{He}{1} $\lambda 6678$ (semi-amplitude $K = 199.9 \pm 6.4$ \kms). 
Dashed line: projected radial velocity of the secondary star according 
to Shahbaz \etal (1999) (semi-amplitude $K = 215.5 \pm 2.4$ \kms).
}
\label{HeI_abs}
\end{figure}

The velocity shifts of the narrow absorption components 
of the \ion{He}{1} lines were consistent with the projected 
radial velocity of the secondary star (Figure~\ref{HeI_abs}), 
suggesting that these lines were due to absorption near the secondary star. 
The apparent small discrepancy between the observed and predicted 
velocities may be due in part to the uncertainty in our measurement of 
the line position, in part to additional absorption by cold gas near the 
rim of the accretion disk or in the accretion stream, or to a non-uniform 
absorption by the photosphere of the secondary star, which was 
strongly irradiated on the side facing the primary.  

We note that \ion{He}{1} $\lambda 7065$ showed a stronger broad absorption 
component briefly on June 8, with EW $\simeq -1.8$ \AA\ and 
FWHM $\simeq 35$ \AA\ $\simeq 1500$ \kms.
It disappeared later on the same night while the hard 
X-ray flux increased, and was not seen during the following nights 
(see Figure~\ref{96jun}).   

Narrow, single-peaked \ion{N}{3} $\lambda 4634$ and 
\ion{N}{3} $\lambda \lambda 4641, 4642$ Bowen fluorescence lines
were observed in 1996 June. 
As these lines were also seen before the hard X-ray 
turned on [see the spectra taken by Hynes \etal (1998b) in 1996 May], 
they were probably uncorrelated with the hard X-ray flux.
By comparing the Bowen line profiles and radial velocity shifts 
with those of the \ion{He}{2} $\lambda 4686$ line, we conclude that 
the Bowen fluorescence lines and the \ion{He}{2} line originated from 
different regions. Judging from the radial velocities shifts 
of the Bowen lines, Soria \etal (1998) suggest that 
they were emitted from a localized region in the 
outer accretion disk, possibly a hot spot in phase with the 
secondary star.

\subsection{Profile of the H$\alpha$ emission component}

Among the emission lines observed in 1996 June, the H$\alpha$ line 
had the highest signal-to-noise ratio.  If we adopt a reddening 
$E(B-V) = 1.2$ (Hynes \etal 1998b), the inferred average flux for 
the H$\alpha$ emission line was 
$F_{\scriptsize {\mbox {H$\alpha$}}} 
\approx 5 \times 10^{-13}$ erg cm$^{-2}$ s$^{-1}$ (cf.\   
$F_{\scriptsize {\mbox {He{\tiny{ II}}}}} \approx 3 
\times 10^{-13}$ erg cm$^{-2}$ s$^{-1}$ 
for the \ion{He}{2} $\lambda 4686$ line).  
Although the EW of the H$\alpha$ line appeared to be correlated 
with the hard X-ray flux (Figure~\ref{batse_EW1}), we have found that 
the velocity profile of the line, with its peak normalized to unity, 
depended only on the binary phase.  

The systematic variation of the H$\alpha$ emission line profiles 
over an orbital cycle can be seen in the sequence of profiles plotted in  
Figures~\ref{Ha_stack1}, \ref{Ha_stack2} and \ref{Ha_stack3}.   
The red-shifted peak was stronger than the blue-shifted peak between 
phase 0.25 and 0.85, with the blue-shifted peak dominating at other phases.  
By considering the averaged, normalized H$\alpha$ line 
profiles over the whole epoch of our observations (see Figure 2 in 
Soria \etal 1998), and the averaged 
profiles obtained between orbital phases 0.80 and 0.92  
and between phases 0.19 and 0.28 (Figure~\ref{Ha_avg}), when 
the highest signal-to-noise ratio was achieved and both peaks had 
comparable strength, we deduce that the FWHM of the line was 
$\simeq 900$ \kms, and the peak-to-peak velocity separation was 
$500 \pm 50$ \kms. These values are similar to those determined for 
the double-peaked \ion{He}{1} and \ion{He}{2} emission lines.

In principle, symmetric double-peaked lines can be produced 
by an accretion disk (Smak 1981; Horne \& Marsh 1986). As the H$\alpha$ line 
profile changed with a period equal to the binary 
period, this eliminates the possibility that the line asymmetry was due to a 
precession of the accretion disk, which should have a period longer than 
the binary period (e.g.\ Kumar 1986, Warner 1995). Neither could it be 
caused by the secondary star eclipsing either side of the disk, because 
our data show asymmetry of the two peaks even when the star was behind 
the accretion disk.  

If the emission were from the irradiatively-heated 
surface of the secondary star, we would expect a stronger blue component 
in the emission line when the star was approaching 
(at phases 0.25 --- 0.75), 
and vice versa. This is inconsistent with the observation that 
the blue peak of the line was stronger at around phase 0. 
Therefore we reject the possibility that the secondary star was 
the dominant source of the emission.     
(See \S5 for more detailed discussion on the H$\alpha$ emission.)

\subsection{1997 June observations}  

The soft X-ray flux in 1997 June was slightly 
lower than in 1996  June, while the hard X-ray flux 
was significantly lower (Figure~\ref{batse_asm}). 
We observed the system on 1997 June 14 and 15. 
No optical photometric observations were carried out during 
this period. Although the conditions were not photometric, 
a comparison with other stars in the field led to an 
estimate of $V \simgt 16.5$ for the source, 
about half a magnitude fainter than in 1996 June. 

The H$\alpha$ emission line had a red-shifted peak around phase 0.5 
and a blue-shifted peak around phase 0, as in 1996 June; 
the signal-to-noise ratio was much lower than in 1996. 
The velocity shifts with respect to the 
center of mass of the binary were $\simeq 255$ \kms for the 
red-shifted peak, and $\simeq -230$ \kms for the blue-shifted peak, 
similar to those measured in 1996. 

For the H$\alpha$ line, EW $=2.5 \pm 0.4$ \AA\ on June 14, and 
$2.7 \pm 0.3$ \AA\ on June 15.  The emission component of the 
H$\beta$ and higher Balmer lines was not detectable. The Balmer 
emission  was therefore weaker than in 1996 June. 
The \ion{He}{2} $\lambda 4686$ line was seen in emission with 
EW $=3 \pm 1$ \AA, as strong as in 1996. The Bowen fluorescence 
\ion{N}{3} lines were also detected.

\section{BALMER EMISSION IN THE HIGH-SOFT STATE}

\subsection{Kinematic interpretation of the H$\alpha$ emission}

\subsubsection{Peak separation}

Figure~\ref{plotpeaks} shows the velocity shift of the 
peaks of the H$\alpha$ emission component detected in the 1996 June and the 
1997 June spectra. Only one peak was present at some 
orbital phases (cf.\ Figures~\ref{Ha_stack1}, \ref{Ha_stack2} 
and \ref{Ha_stack3}). The velocity separation of the peaks 
($\simeq 500$ \kms) was larger than the radial velocity amplitude of the 
secondary star. This implies that the emission region 
was located within the orbit of the secondary star, thus ruling 
out the possibility of emission from the surface of the secondary 
(see also \S4.3) or from a circumbinary disk.   

According to Orosz \& Bailyn (1997), the projected radial velocity 
semi-amplitude of the secondary star in GRO~J1655$-$40 is 
$K_2=228.2 \pm 2.2$ \kms, and the masses of the two components  
are $M_1 \simeq 7 M_{\odot}$ and $M_2 \simeq 2.3 M_{\odot}$.
If we assume a thin, non-circular, Keplerian accretion disk 
truncated at the tidal radius, as described 
in Paczy\'{n}ski (1977), then the disk emission lines would have peaks 
with observed radial velocities at each phase given by the dash-dotted 
lines in Figure~\ref{plotpeaks}. The tidal truncation radius is approximately 
given by $R_{\rm d} = 0.60 \, a /(1+q)$ for $0.03 < q < 1$ (Warner 1995), 
where $a$ is the separation between the centers of mass of the 
binary components. The peak-to-peak velocity separation 
averaged over an orbital cycle would be $\simeq 820$ \kms. 
Phillips \etal (1999) derived lower values for the radial velocity 
semi-amplitude of the secondary star
and for the masses of the binary components. They obtained 
$192 < K_2 < 214$ \kms, 
$4.1 < M_1 < 6.6 M_{\odot}$ and $1.4 < M_2 < 2.2 M_{\odot}$ (90 per cent 
confidence limit). For a fixed mass ratio $q=1/3$, this implies that 
the peak-to-peak velocity separation averaged over an orbital cycle, 
expected from a tidally-truncated Keplerian disk, would 
be $690 < \Delta V < 760$ \kms. Peak separations $\simeq 770$ \kms 
would instead be expected if we adopt the parameters determined 
by Shahbaz \etal (1999), who inferred a velocity semi-amplitude 
$K_2=215.5 \pm 2.4$ \kms for the companion star.

The velocity separations deduced from 
our 1996 June and 1997 June data are lower than the predicted values, 
especially for the red-shifted peak at phases around 0.5 and 
the blue-shifted peak at phases around 0. Therefore, the double-peaked 
profile of the H$\alpha$ emission component cannot be explained by 
a conventional thin Keplerian accretion disk truncated at the tidal radius.  
We note that the peak-to-peak velocity separations of \ion{He}{2} 
$\lambda 4686$ and \ion{He}{1} $\lambda 6678$ were 
also too small to be consistent with a tidally-truncated Keplerian disk.

Remarkably low rotational velocities, inconsistent with Keplerian 
disks truncated at their tidal radii, have also been inferred from the 
double-peaked H$\alpha$ emission line profiles observed from A0620$-$00 
and GS~1124$-$68 (Orosz \etal 1994).
A possible explanation for the low rotational velocities inferred 
for the outer edge of the accretion disks in these three systems
is that the disks extended slightly beyond their tidal radii. 
(cf.\ Whitehurst 1988).

\subsubsection{A schematic model}

We have found that the observed variations in the H$\alpha$
line profile can be explained by the model shown in 
Figure~\ref{sectors96}, with two separate 
emission regions on the accretion disk.  
The model can reproduce the low radial velocities of the 
peaks and their behavior at various orbital phases. For 
example, only the red-shifted component would be visible around phase 0.5 and 
only the blue-shifted component around phase 1, while both peaks 
would be visible around phases 0.25 and 0.75.

Non-axisymmetric patterns of emission similar to our model 
were proposed by Steeghs, Harlaftis \& Horne (1997) to explain the 
Balmer and helium emission line profiles observed from the dwarf nova 
IP Peg, and by Neustroev \& Borisov (1998) to explain the Balmer 
line profiles observed from the dwarf nova U Gem. In both cases, 
they were interpreted as evidence of a two-armed spiral density wave 
or shock in the accretion disk, induced by the tidal interaction with 
the companion star. In the case of IP Peg, the observed rotational 
velocities also suggested that the disk could extend beyond its tidal 
radius (Steeghs \etal 1997).

It has been suggested that accretion stream overflows, hot spots, and 
uneven illumination of the accretion disk from the central X-ray 
source can also produce an anisotropic emissivity. However, it is beyond the 
scope of the present paper to explore in details why the 
Balmer emission regions might assume the geometrical configuration 
shown in our model.   

\begin{figure}
\epsfxsize=145mm\epsfbox{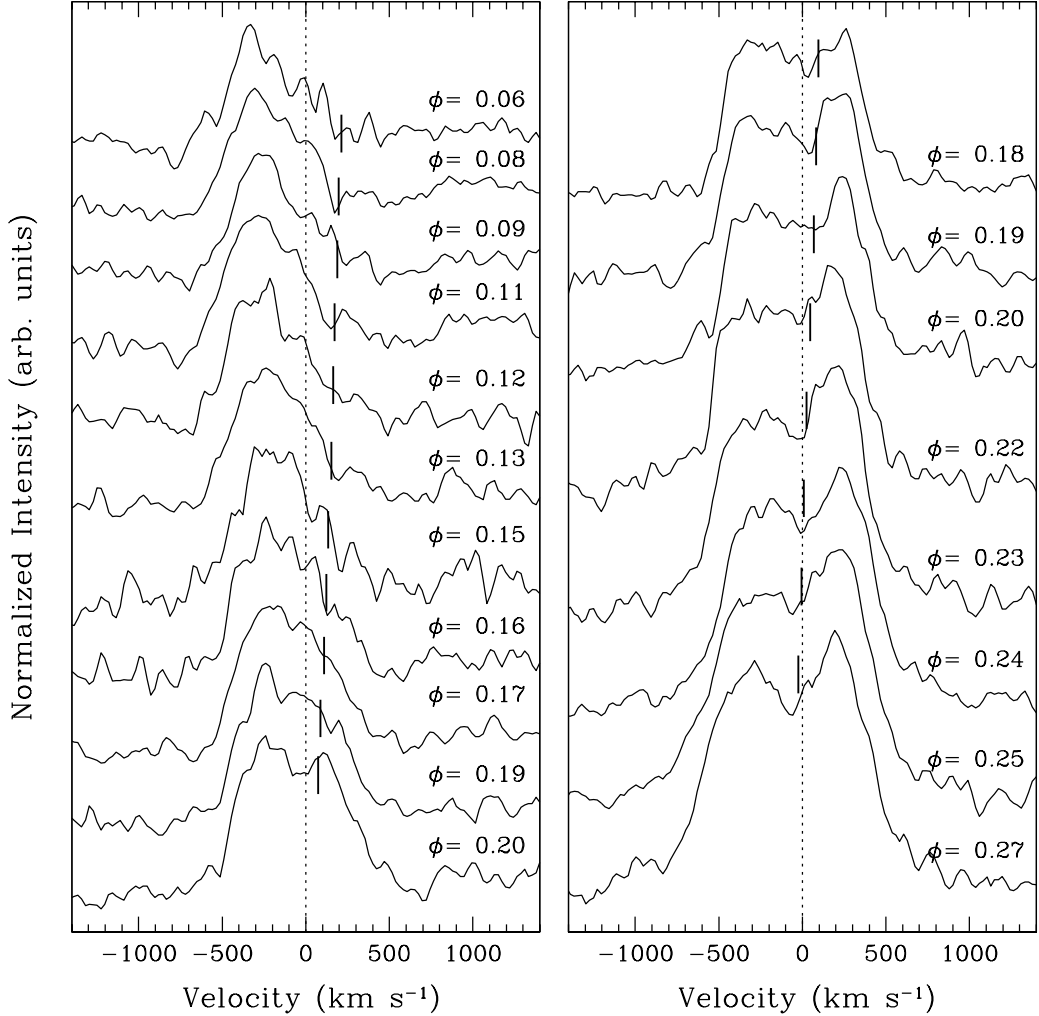} 
\caption{H$\alpha$ emission line profiles observed on 1996 June 8 
(left panel) at spectroscopic phases $0.06 < \phi < 0.20$, 
and on June 11 (right panel), at phases 
$0.18 < \phi < 0.27$.   The velocity 
zeropoint (dotted line) is the systemic velocity 
($\gamma = -142.4 \pm 1.6$ \kms).  
The projected radial velocities of the secondary  
star with respect to the systemic velocity are also shown next to each 
spectrum (short solid lines). Note that when the secondary star was 
receding from us, the largest contribution to the H$\alpha$ emission 
was coming from approaching gas. Both the red and the blue peak 
were visible when the secondary star was near superior conjunction. }
\label{Ha_stack1}
\end{figure}

\begin{figure}
\epsfxsize=145mm\epsfbox{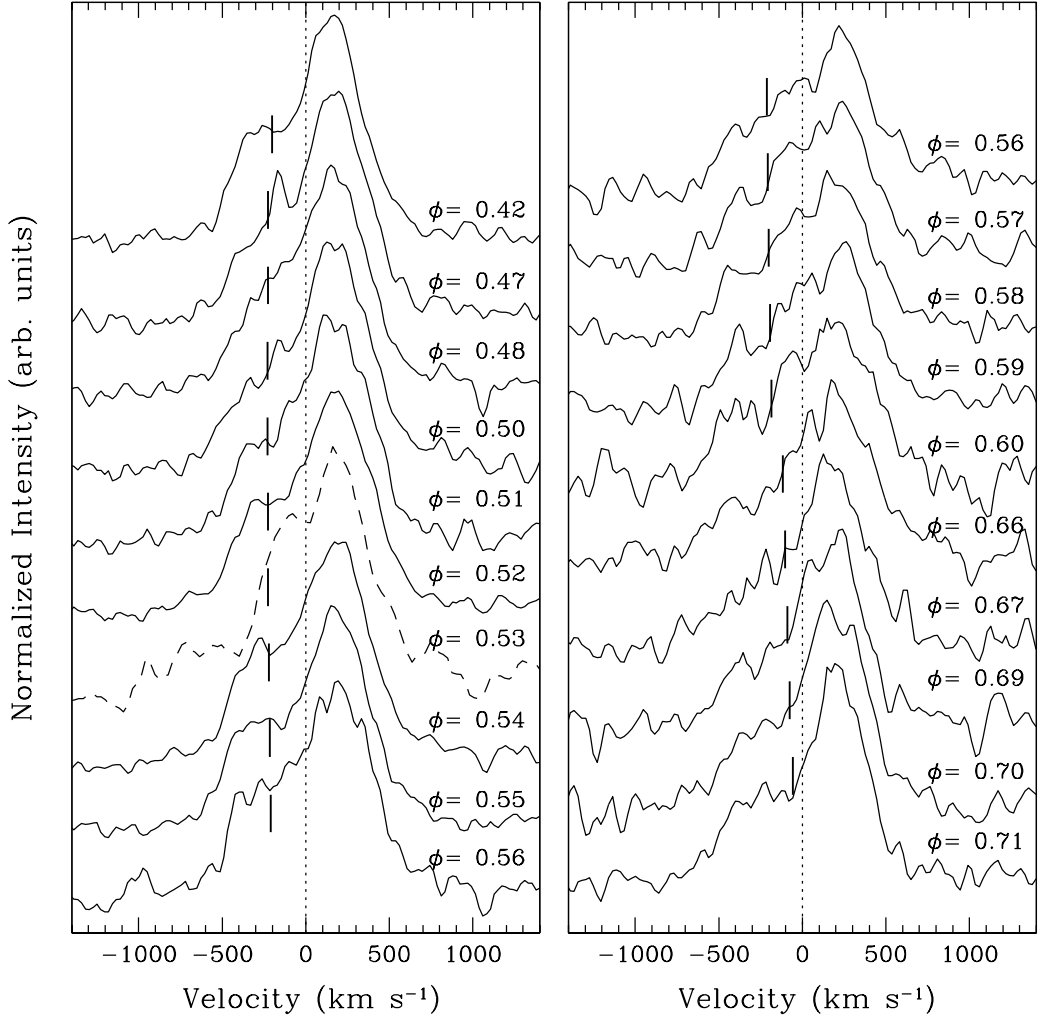}  
\caption{As in Fig.~\protect\ref{Ha_stack1}, for the 1996 June 9 observations 
(left panel), at spectroscopic phases $0.42 < \phi < 0.56$, and the June 12 
observations (right panel), at phases $0.56 < \phi < 0.71$. The dashed 
profile in the left panel was obtained on 1996 June 17. Note that while 
the secondary star was approaching us, most of the emission was coming from 
receding gas. A narrow absorption feature, with a velocity consistent with 
the velocity of the companion star, is visible for 
$0.5 \simlt \phi \simlt 0.7$, superimposed on the emission 
profile.}
\label{Ha_stack2}
\end{figure}

\begin{figure}
\epsfxsize=145mm\epsfbox{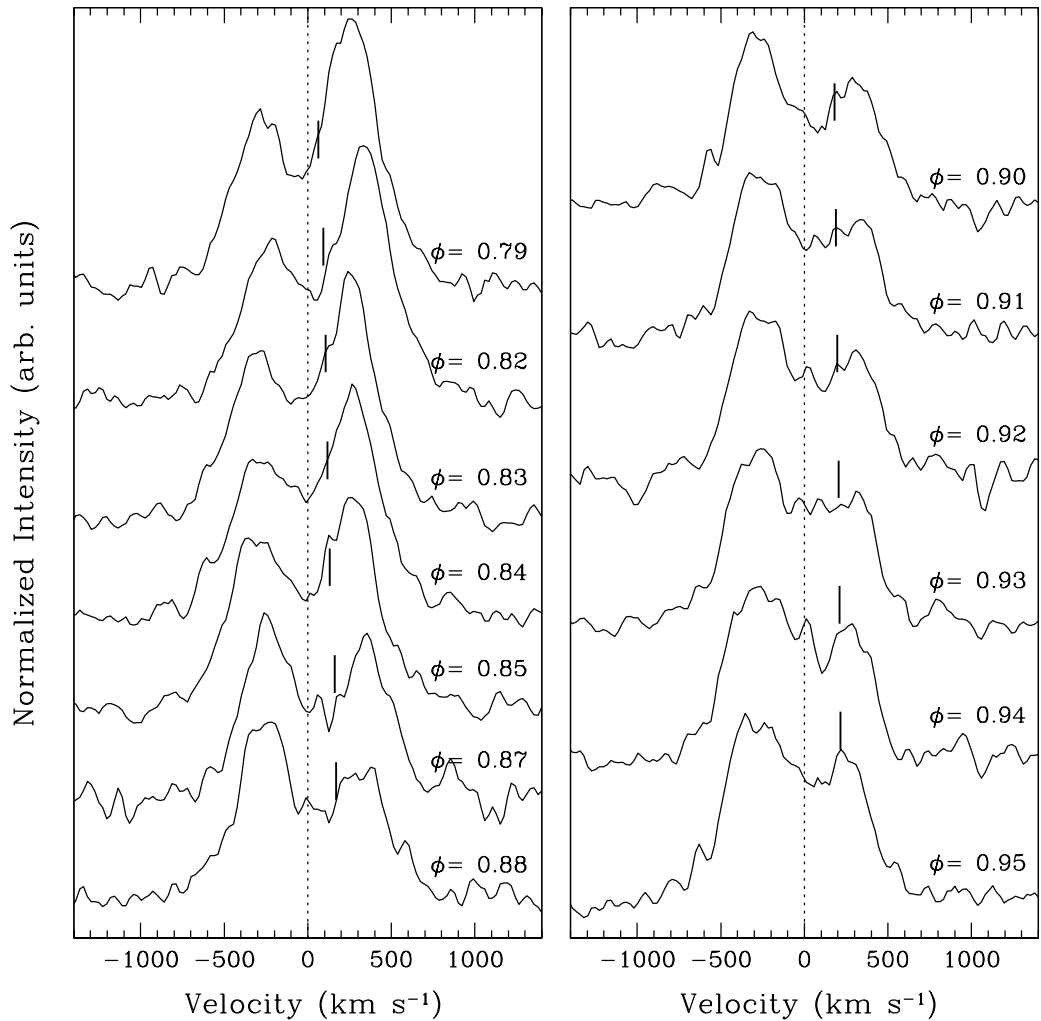}  
\caption{As in Fig.~\protect\ref{Ha_stack1}, for the 1996 June 10 observations, 
at spectroscopic phases $0.79 < \phi < 0.88$
(left panel) and at phases $0.90 < \phi < 0.95$
(right panel).  }
\label{Ha_stack3}
\end{figure}

\begin{figure}
\epsfxsize=145mm\epsfbox{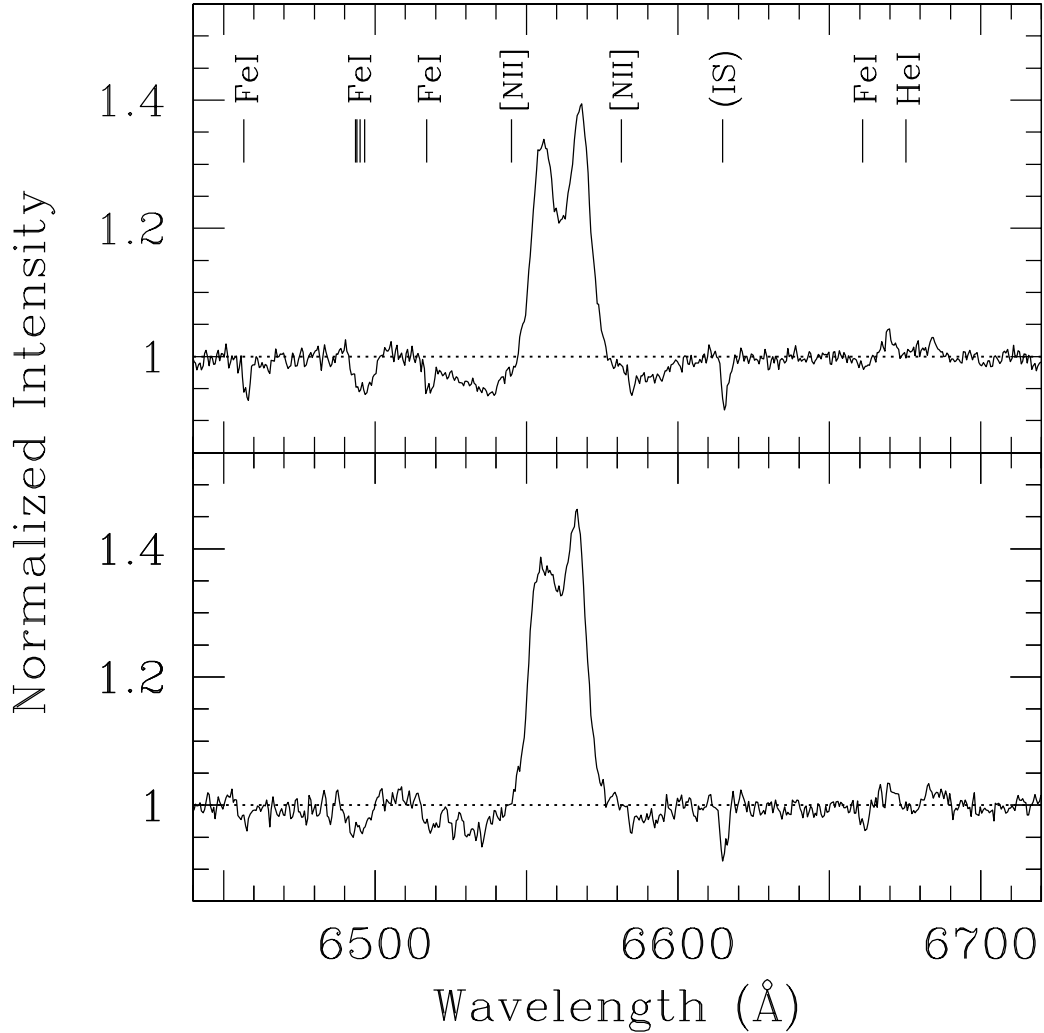}  
\caption{Top panel: normalized averaged spectrum in the H$\alpha$ region, 
obtained on 1996 June 10, at spectroscopic phases $0.80 < \phi < 0.92$. 
Bottom panel: normalized averaged spectrum obtained on 1996 June 11, 
at phases $0.19 < \phi < 0.28$.
Wavelengths are vacuum heliocentric. Double-peaked emission was observed 
at H$\alpha$ and \protect\ion{He}{1} $\lambda 6678$. Weak 
[\protect\ion{N}{2}] $\lambda 6548$ and [\protect\ion{N}{2}] $\lambda 6584$ emission lines 
were probably blended with the wings of H$\alpha$.
Narrow \protect\ion{Fe}{1} absorption lines, probably from the secondary 
star, were also seen. (They appear broadened here because 
the spectra have been averaged over a large phase interval; in particular, 
they appear double-peaked in the top panel because the spectra at 
$\phi \simeq 0.80$ and $\phi \simeq 0.90$ give the largest contribution 
to the average spectrum.)}
\label{Ha_avg}
\end{figure}

\begin{figure}
\epsfxsize=135mm\epsfbox{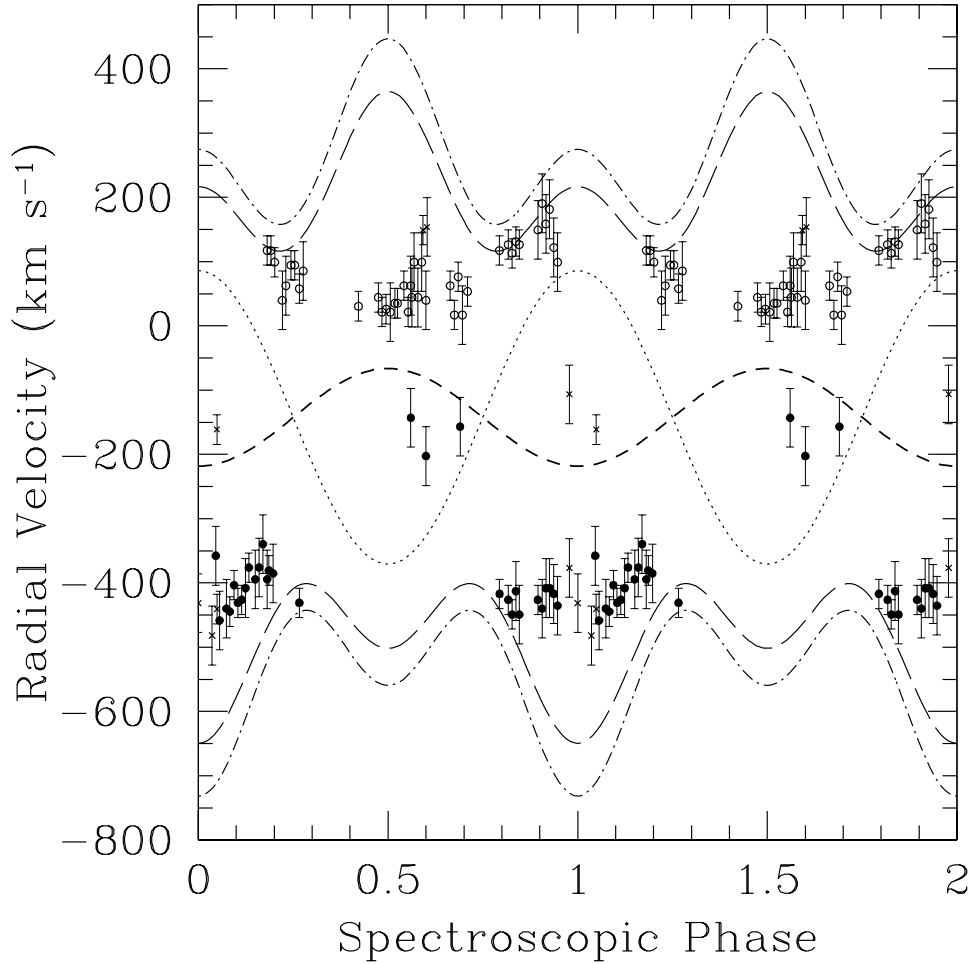}  
\caption{Velocities of the peaks detected in the H$\alpha$ emission 
line profiles, as a function of spectroscopic binary phase, compared with 
other characteristic velocities of the system. The zeropoint in the 
velocity scale is referred to the heliocentric system; the systemic 
velocity of the centre of mass is taken as $-142.4 \pm 1.6$ \kms (Table 1).
Open and filled circles indicate the velocity shifts of the 
red and blue peaks respectively, detected in the 
1996 June spectra. Crosses indicate the peaks detected in the 
1997 June spectra. 
The short-dashed line is the projected radial velocity of the primary star, 
and the dotted line is the projected radial velocity of the secondary 
star, as determined by Orosz \& Bailyn (1997). (Smaller velocities 
for the two components were inferred by Phillips \etal 1999 and 
Shahbaz \etal 1999). 
The dash-dotted lines are the projected velocities of the two
edges of a non-circular Keplerian disk truncated at the tidal radius, as 
calculated by Paczy\'{n}ski (1977), for the system parameters determined 
by Orosz \& Bailyn (1997); the long-dashed lines are calculated
for the system parameters determined by Phillips \etal (1999). 
The velocity corresponding to the parameters inferred by Shahbaz \etal 
1999 would be intermediate between the two.}
\label{plotpeaks}
\end{figure}

\begin{figure}
\epsfxsize=145mm\epsfbox{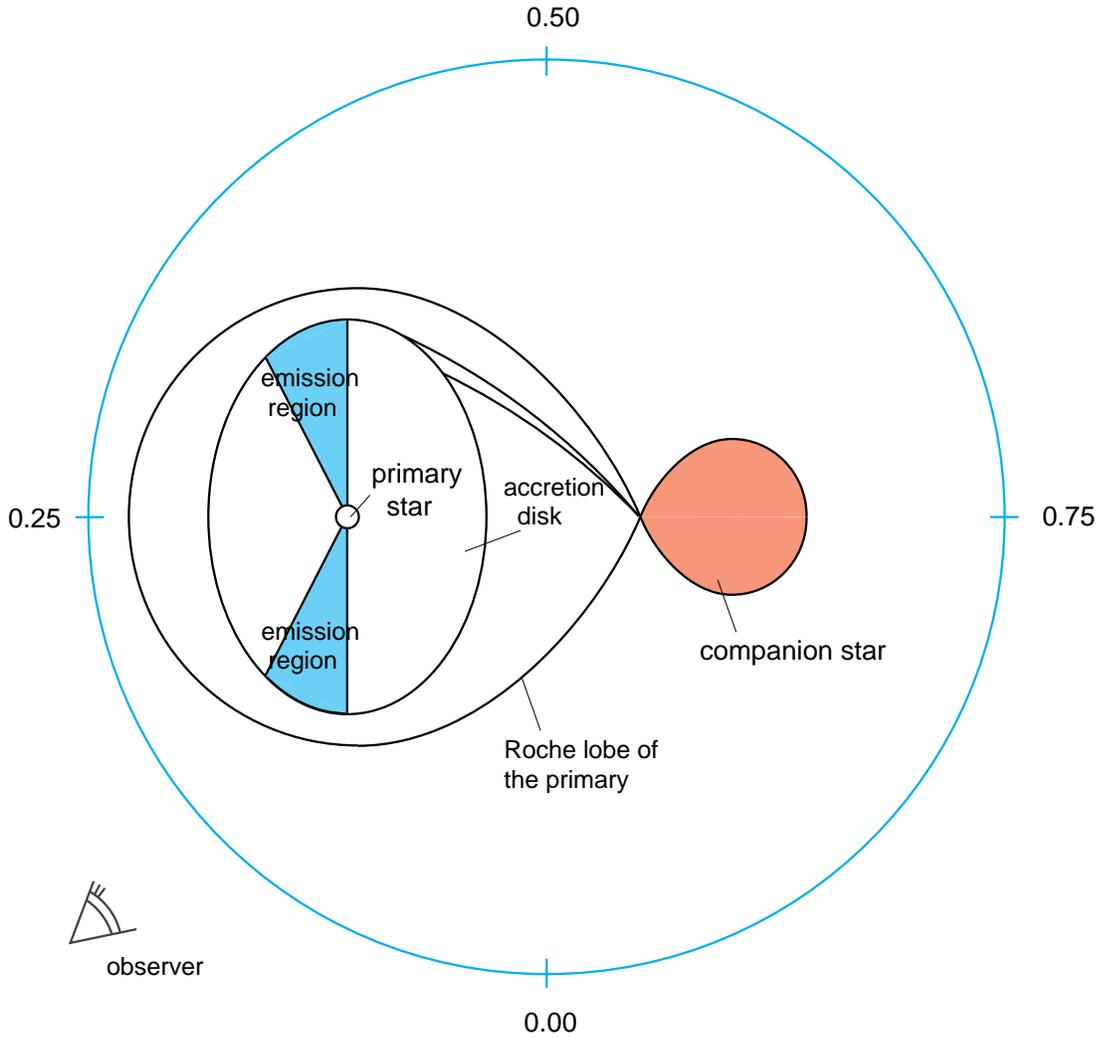}  
\caption{Schematic elliptical disk model that can reproduce the  
  low velocity separations of the H$\alpha$ emission-line peaks, 
  and the line profile variations over an orbital period 
  as observed in 1996 June and 1997 June. In the model, H$\alpha$ 
  would be emitted mostly from the shaded regions.
  We do see both emitting regions at all phases, but it can be noticed 
  from the figure that at some phases both regions are blue-shifted 
  (around phase 0) or red-shifted (around phase 0.5).}
\label{sectors96}
\end{figure}

Furthermore, a narrow $H\alpha$ absorption component from the companion 
star is also likely to be present, as suggested by the detection 
of other stellar absorption features from \ion{He}{1} and \ion{Fe}{1}. 
It can be noticed from the radial velocity curve of the companion star 
that the presence of a stellar absorption line 
superimposed on the disk emission would also be qualitatively consistent 
with the observed behavior of the line profile at various orbital phases, 
if absorption and emission have comparable strength 
(Figures~\ref{Ha_stack1}, \ref{Ha_stack2} and \ref{Ha_stack3}).
However, we argue that the effect of anisotropic disk emission 
is more significant than the effect of a stellar absorption line, 
and can better explain the velocity shifts of the emission peaks 
observed over an orbital phase.

One might expect the H$\alpha$ emission line profiles to be roughly 
symmetric at superior and inferior conjunction; however, 
the H$\alpha$ line in general showed a red-shifted peak 
stronger than the blue-shifted peak, and a blue wing more extended than 
the red wing (see Figure~\ref{Ha_stack1} and Figure~\ref{Ha_stack3}). 
The asymmetry of the profiles may indicate the presence of both opacity 
and kinematic effects, and may be evidence of a thin disk-wind. 
It is worth noting that P-Cygni profiles were observed 
in some UV resonance lines from the system (Hynes \etal 1998b).

\subsection{Formation of absorption and emission lines}

\subsubsection{Broad absorption lines}

The H$\alpha$ and H$\beta$ lines in the spectra obtained during 
the 1996/1997 outburst all show broad (FWHM $\approx 3000$ \kms), 
shallow absorption components. 
Broad absorption components are also seen in the 1994 August -- 
September spectra, when the system was in outburst 
(see \S6.2), but not in the 1996 April  
spectra, when the system was in quiescence. 

Absorption lines can be produced in an accretion disk if the disk 
is optically thick and its temperature decreases with height above the 
central plane. The absorption features seen in our spectra are 
broader than the emission cores: this suggests that they were probably 
formed in the inner part of an optically thick disk, where viscous 
heating from the central plane dominates over external irradiative heating.

Broad, shallow absorption features at the Balmer lines have been seen 
in the spectra of other BHCs in outburst, such as A0620$-$00 
(Whelan \etal 1977), GS~1124$-$68 (Della Valle, Masetti \& Bianchini 1998) 
and, most prominently, GRO~J0422$+$32 (Casares \etal 1995). Similar 
features are also often present in the spectra 
of UX UMa stars (e.g.\ Warner 1995) and in the outburst spectra of 
Dwarf Novae (Robinson, Marsh \& Smak 1993).

\subsubsection{Double-peaked emission lines}

Double-peaked lines can be emitted from a hot temperature-inversion layer 
on the X-ray irradiated surface of the accretion disk.
Provided that the spectrum is sufficiently soft, 
X-rays can be absorbed at a small depth, forming a thin temperature-inversion 
layer but leaving most of the vertical structure of the disk 
undisturbed (see Tuchman, Mineshige \& Wheeler 1990; Wu \etal 1999).
 
The strong, double-peaked \ion{He}{2} $\lambda 4686$ line 
seen during the high-soft state in 1996--1997 was probably emitted 
via radiative recombination in this thin, hot layer, 
at temperatures $\sim 10^5$ K.

While the \ion{He}{2} $\lambda 4686$ was detected throughout the 1996 
outburst, Balmer emission was seen only after the hard X-rays turned on.
Balmer lines are likely to be emitted at lower temperatures ($\sim 10^4$ K).
Moreover, the H$\alpha$ double-peaked profile is strongly asymmetric and 
phase-dependent (\S5.1.2), while the \ion{He}{2} $\lambda 4686$ profile is 
approximately symmetric at {\em all} phases.
For all these reasons we suggest that \ion{He}{2} $\lambda 4686$ and 
the Balmer lines were emitted from different regions: the Balmer lines 
probably originated from a deeper, denser layer in the disk where matter 
was heated by harder X-ray photons.

\section{ONSET OF A HARD X-RAY OUTBURST (1994 AUGUST -- SEPTEMBER)}   

\subsection{Overview} 

GRO J1655$-$40 was active in the radio and hard X-ray energy bands 
in 1994 August -- September. The 843 MHz radio flux density, 
measured by the MOST, was declining after a large outburst that 
reached about 8 Jy in early August (Wu \& Hunstead 1997).  
On 14 September another radio outburst started. It was weaker 
than the previous outburst, with a peak flux density of about 2 Jy. 

The hard X-ray flux recorded by BATSE (Harmon \etal 1995) was below 
the 0.7 photon cm$^{-2}$ s$^{-1}$ level 
during the period August 15 -- September 4. A sharp rise in the X-ray 
flux occurred on September 5, and the flux stayed at the 
2 photon cm$^{-2}$ s$^{-1}$ level for about 10 days. It then declined 
abruptly, apparently in coincidence with the rise in 
the 843 MHz radio luminosity around September 14 
(see Fig.~1 in Wu \& Hunstead 1997). 

The span of our observations covered the transition phase around the 
hard X-ray rapid increase. Good signal-to-noise spectra, 
with a resolution of 1.3 \AA\ FWHM, in the spectral regions 
centered at H$\alpha$ and at \ion{He}{2}/ H$\beta$, 
were obtained on August 30 -- September 4, before the rise in the 
hard X-ray flux. The sampling of binary phase was roughly uniform 
over the six-night observations. Two lower-resolution spectra 
in the H$\alpha$ and \ion{He}{2}/ H$\beta$ regions 
were taken on 1994 September 6, after the rise in the X-ray flux, and  
showed a dramatic change compared with the August 30--September 4 spectra.

\subsection{General spectral features before September 6} 

\begin{figure}  
\epsfxsize=145mm\epsfbox{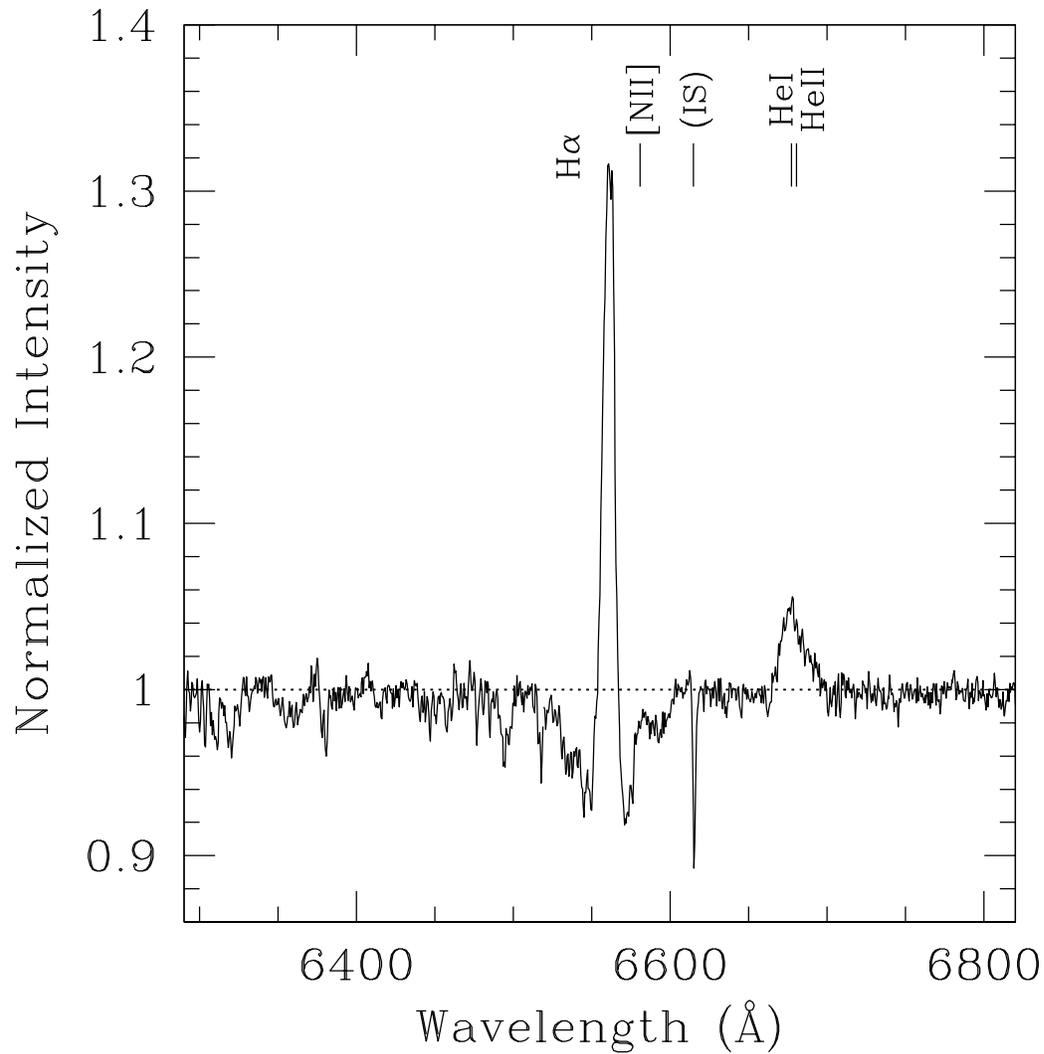}
\caption{Averaged red spectrum obtained 
by combining the spectra taken from 1994 August 30 to September 4 
with the 3.9~m Anglo-Australian Telescope at Siding Spring Observatory. 
Wavelengths are vacuum heliocentric and the intensity is 
normalized to the continuum. The most prominent features in the 
spectrum are marked.}
\label{total94_ha}
\end{figure}

In Figure~\ref{total94_ha} we show the red spectrum, 
centered at H$\alpha$, averaged over the six nights from August 30 to 
September 4. As in the spectra obtained during the 1996 -- 1997 
high-soft state, we notice a broad absorption component partly filled by 
narrow H$\alpha$ emission. The average EW of the absorption 
component is $-6 \pm 1$ \AA,  and its average FWHM 
$60 \pm 10$ \AA\ ($\simeq 2700 \pm 500$ \kms). The 
narrow emission component has an average EW $= 5.3 \pm 0.2$ \AA\  
and an average FWHM $= 450 \pm 20$ \kms.

\begin{figure}  
\epsfxsize=145mm\epsfbox{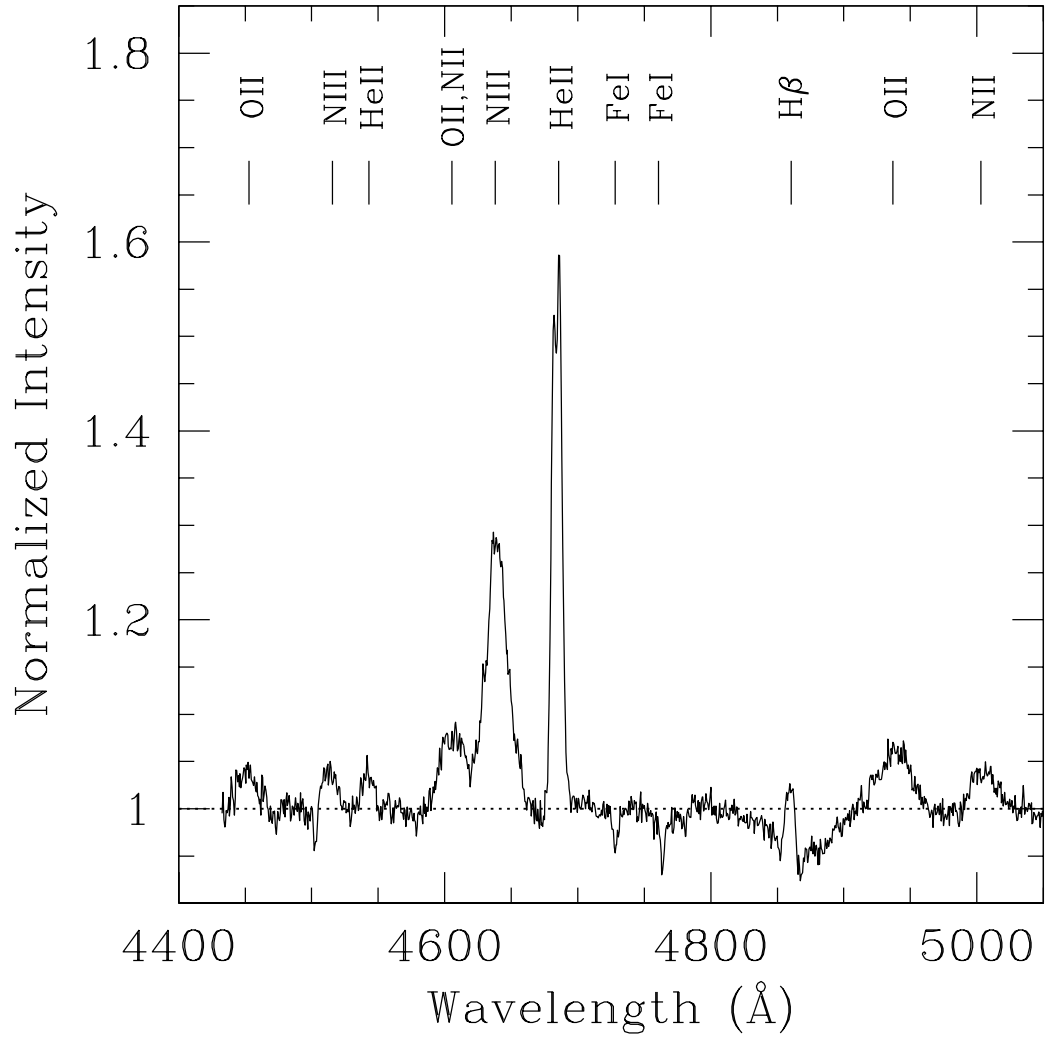}  
\caption{Averaged blue spectrum obtained 
by combining the spectra taken from 1994 August 30 to September 4. 
Wavelengths are vacuum heliocentric and the intensity is 
normalized to the continuum. The most prominent features in the 
spectrum are marked. We notice the presence of broad absorption, 
broad (flat-topped) emission and narrow emission lines.}
\label{total94_hb1}
\end{figure}

Figure~\ref{total94_hb1} shows the averaged blue spectrum, 
centered at \ion{He}{2}/ H$\beta$. The broad absorption component 
of H$\beta$ has an EW $= -4 \pm 1$ \AA\ 
and an FWHM $= 50 \pm 10$ \AA\ ($\simeq 3000 \pm 600$ \kms). 
A narrow emission component is superimposed on the broad absorption 
trough, with EW $= 1.1 \pm 0.1$ \AA\ and FWHM $= 550 \pm 30$ \kms.

\ion{He}{2} $\lambda 4686$ appears in the spectrum as a strong narrow 
emission line (Figure~\ref{total94_hb1}). 
The EW of the line is $ 5.2 \pm 0.2$ \AA\ and its FWHM is $540 \pm 20$ \kms. 
Weaker emission is detected at \ion{He}{2} $\lambda 4542$, with 
EW $= 0.5 \pm 0.2$ \AA\ and FWHM $= 800 \pm 50$ \kms.

\begin{figure}  
\epsfxsize=145mm\epsfbox{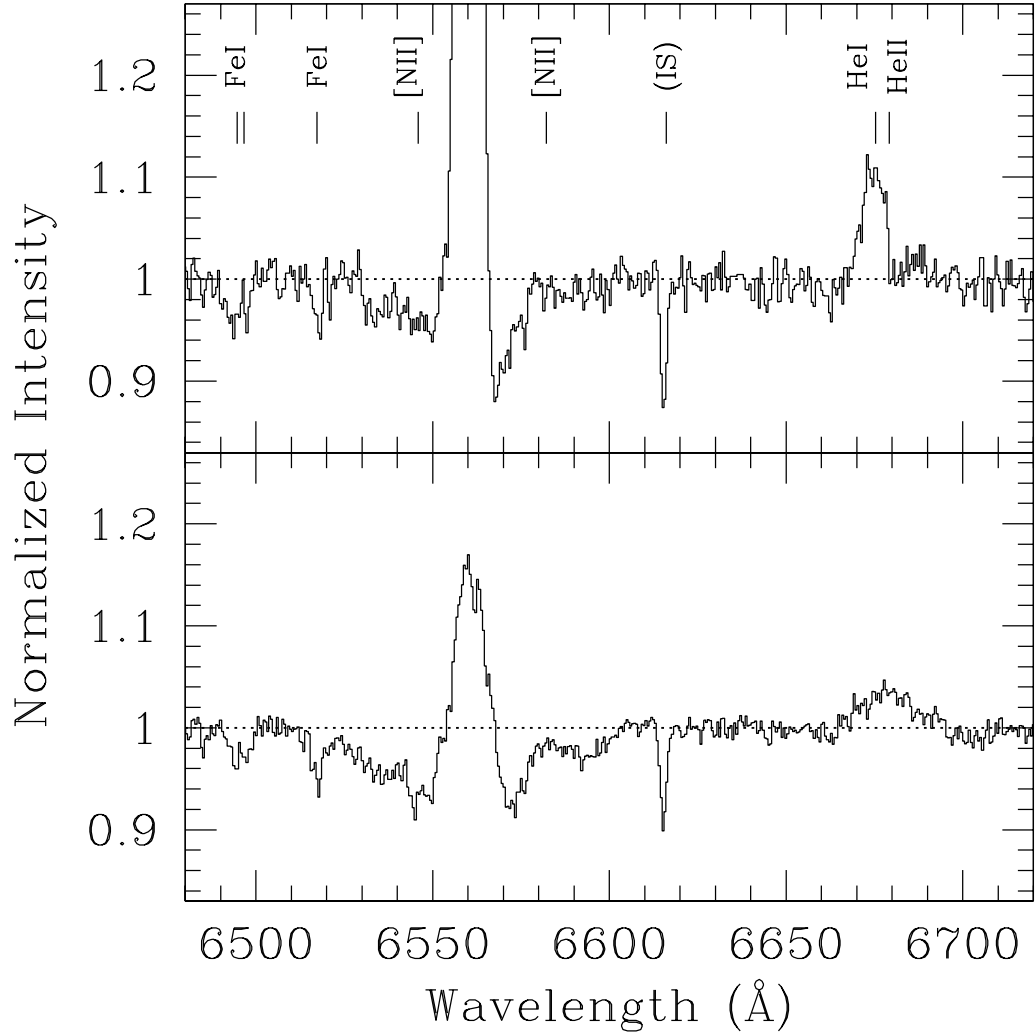} 
\caption{Top panel: expanded view of the H$\alpha$ region in the normalized 
spectrum from August 30. Bottom panel: expanded view of the same region in the 
normalized, combined spectrum from August 31 and September 1.
The narrow emission components at H$\alpha$ and \protect\ion{He}{1} $\lambda 6678$ 
were much stronger on August 30 than on the following two days.}
\label{n_narrow94}
\end{figure}

A broad emission component (FWHM $= 950 \pm 100$ \kms) was  
detected on each night at \ion{He}{1} $\lambda 6678$. An additional narrower 
component (FWHM $= 450 \pm 50$ \kms) was detected on some nights; it was 
particularly strong on August 30 (cf.\ Figures~\ref{n_narrow94}). We note 
that this line might be contaminated by \ion{He}{2} $\lambda 6683$ emission.

The spectra show a broad, flat-topped line at about $5005$ \AA. It 
cannot be attributed to [\ion{O}{3}] 
$\lambda 5007$ both because the wavelengths are discrepant, and 
because we do not see any emission from [\ion{O}{3}] $\lambda 4959$, 
whose strength should be one-third of that of 
[\ion{O}{3}] $\lambda 5007$. 
(The only forbidden lines possibly detected in our spectra obtained 
between 1994 August 30 and September 4 are [\ion{N}{2}] $\lambda 6548$, 
blended with the much stronger blue wing of H$\alpha$, and 
[\ion{N}{2}] $\lambda 6589$; see Figures~\ref{n_narrow94}).   
We identify the line at about $5005$ \AA\ as \ion{N}{2} $\lambda 5005$. 

Another strong 
low-ionization metal line detected in the spectrum is the blend 
\ion{O}{2} $\lambda \lambda 4941,4943$. Other broad emission lines 
detected from \ion{N}{2} and \ion{O}{2} are listed in Table 3.
We do not include a weak emission line at $6375$ \AA\ in Table 3, 
as it was seen only on the first two nights (EW $= 0.7 \pm 0.1$);  
its wavelength is consistent with that of an \ion{Fe}{10} line. 
Bowen \ion{N}{3} emission lines were also observed, broader than in 
1996 June, and with a radial velocity curve consistent with that of the 
primary rather than the secondary.

\subsection{Line classification}

The FWHMs of the optical lines observed before 1994 September 6 span 
a large range of values, from $\simlt 400$ \kms to $\approx 3000$ \kms.
The broadest lines are those seen in absorption at H$\alpha$ and H$\beta$. 
We notice that a broad absorption trough 
at H$\beta$ was also detected in the 1995 March outburst 
by Bianchini \etal (1997), and during the 1996 May--June high-soft state. 
The physical interpretation of the broad Balmer absorption lines is 
probably the same as discussed in \S5.2.1. 
Henceforth, we will focus only on the emission lines.

For the orbital parameters of GRO~J1655$-$40, the projected rotational 
velocity of the outer rim of a thin, Keplerian accretion disk, 
truncated at or slightly beyond its tidal radius, is $\simgt 350$ \kms.
Therefore, any lines with an FWHM $< 700$ \kms cannot come from 
a thin Keplerian disk.
We classify the emission lines observed 
before 1994 September 6 as ``broad'' and ``narrow'' 
(Table 3), according to whether their FWHM was 
larger or smaller than the minimum value of FWHM expected for disk 
emission. 

The broad emission lines were usually flat-topped, while 
the narrow emission lines were either single-peaked or had a hint of 
a double-peaked profile but with very low velocity separation.
In some cases the distinction is blurred, especially 
for the weak lines. However, in most situations the two kinds 
of line profiles can be discerned.
We notice that both the EW and the FWHM of the narrow lines 
changed significantly from night to night, while EW and FWHM of 
the broad lines were more stable. 

Our phenomenological classification of broad 
and narrow emission lines in GRO J1655$-$40 is similar to 
the classification of emission lines in AGNs. For example, 
a study of 123 high-luminosity AGNs (Wills \etal 1993) has 
shown that the profile of the \ion{C}{4} $\lambda 1549$ emission 
line consists of a ``broad'' base and a ``narrow'' core, which 
are emitted from distinct regions. The FWHM of the line is determined 
by the core/base ratio: when the core is dominant, the line will appear 
narrow and sharply peaked; and when the base is dominant the line 
will be broad and flat-topped (cf.\ Fig.\ 2 in Wills \etal 1993).

\subsubsection{Origin of the ``broad'' emission lines}

The low-ionization metal lines from \ion{N}{2} and \ion{O}{2} detected 
in our spectra are examples of broad emission lines; their FWHMs are 
consistent with a disk origin. Unlike the disk emission lines seen in 
1996 June, 
their profiles were generally flat-topped rather than double-peaked.

Flat-topped lines can be emitted in 
a wind from the surface of the accretion disk. Murray \& Chiang (1997) 
showed that if the lines have substantial optical depth ($\tau_l \simgt 1$), 
the contribution of the wind from the front and back sectors of the 
disk (where the projected radial velocities are lower) is enhanced 
relative to the contribution of the wind from the sides of the disk 
(where the projected radial velocities 
are higher), thus smearing out the two peaks in the line profile.

From the fact that we detected only low-ionization broad lines 
from \ion{He}{1}, \ion{O}{2}, \ion{N}{2} and \ion{N}{3}, we infer 
an ionization parameter 
$U \equiv L_{\rm x}/n_{\scriptsize {\mbox {H}}}r^2 \approx 10$ 
in the line-emitting disk wind. The corresponding temperature would be 
$\sim 1$ -- $2 \times 10^4$ K if the gas was collisionally ionized 
(Hatchett, Buff \& McCray 1976); the temperature would be lower 
if the line-emitting gas was photoionized.

The possibility of an outflow or disk wind suggested by the broad optical 
emission line profiles observed in the 1994 X-ray 
outburst was already discussed in Hunstead \etal (1997), who 
compared the spectrum of GRO J1655$-$40 with the spectra of 
WN6-8 Wolf-Rayet stars. In spite of the similarities, there are several 
differences between the two cases. Firstly, high-ionization emission lines  
usually found in WR stars, e.g.\ from \ion{N}{4} and \ion{N}{5}, 
were not observed, indicating a lower temperature for the outflow 
in GRO~J1655$-$40.
Secondly, no forbidden [\ion{O}{3}] lines were observed, 
implying a higher density. 
Thirdly, the broad, flat-topped profiles are 
better explained by a disk-wind model rather than 
by a spherical outflow as in typical Wolf-Rayet stars.

\subsubsection{Kinematics of the ``narrow'' emission lines}

\begin{figure}  
\epsfxsize=145mm\epsfbox{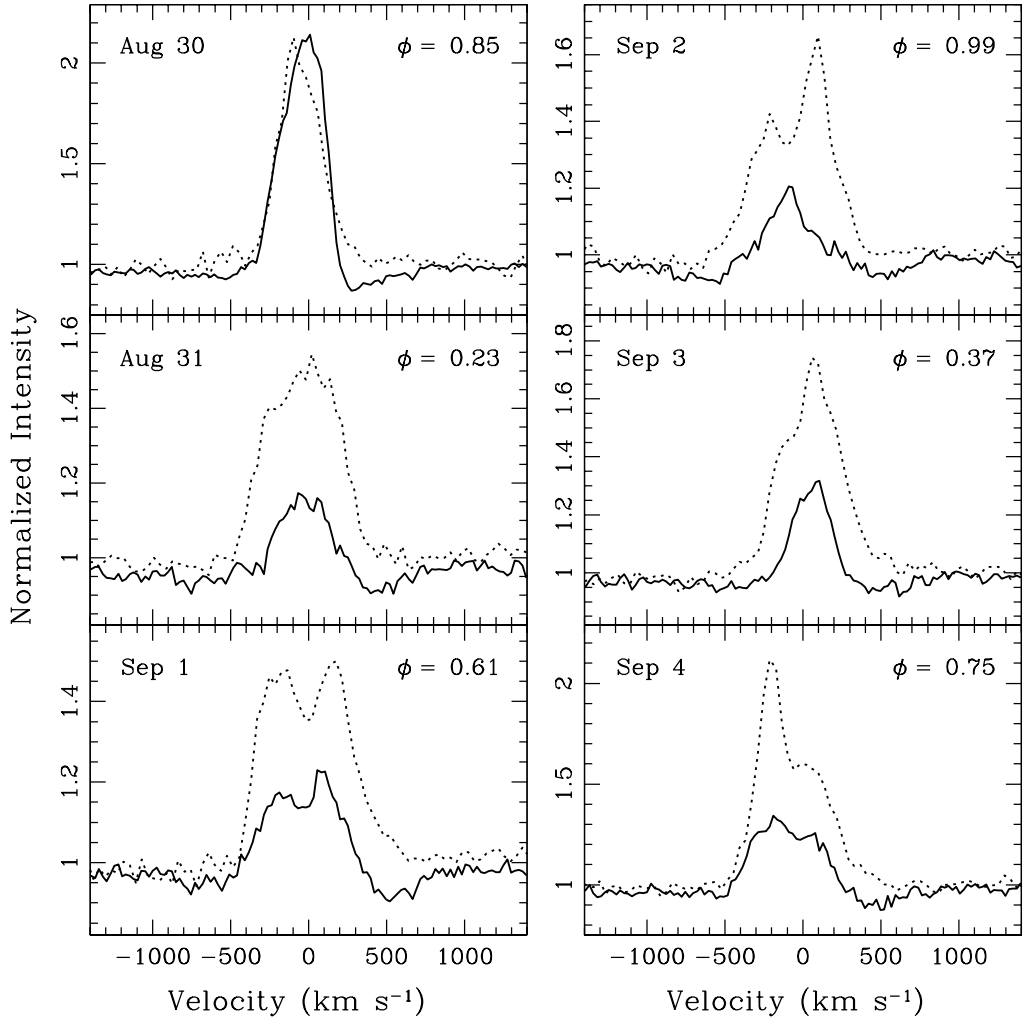} 
\caption{H$\alpha$ (solid curves) and \protect\ion{He}{2} $\lambda 4686$ (dashed 
curves) line profiles observed between 1994 August 30 and September 4. 
A red and a blue spectrum were taken in succession on each night. 
The phase indicated in each diagram refers to the observations of H$\alpha$; 
the phases of \protect\ion{He}{2} $\lambda 4686$ are $\phi_{H\alpha} + 0.01$.
\protect\ion{He}{2} $\lambda 4686$ appeared generally broader than H$\alpha$, 
an indication that the line was emitted at smaller radii or closer 
to the disk plane. Both lines showed transitions between a 
single-peaked and a double-peaked kind of profile. The velocity 
zeropoint is here the systemic velocity, $-142.4$ \kms. On average, 
both lines were blueshifted with respect to the systemic velocity; see \S6.5.}
\label{n_stack94}
\end{figure}

The narrow emission lines showed large night-to-night changes in their 
EWs and line profiles. In particular, they showed transitions between 
narrow, single-peaked and slightly broader, double-peaked profiles 
over the time-span of our observations. We notice that 
\ion{He}{2} $\lambda 4686$ line was generally 
broader and more clearly double-peaked than H$\alpha$ (Figure~\ref{n_stack94}; 
cf.\ also the double-peaked profile in the combined spectrum shown in 
Figure~\ref{total94_hb1}); its peak-to-peak velocity separation 
was however always $< 400$ \kms.

Some lines (most notably \ion{He}{1} $\lambda 6678$) were probably 
emitted from both the broad- and the narrow-line regions, 
with the intensity ratio determined by the relative contribution of the two 
components, i.e.\ as in the ``base'' vs ``core'' classification scheme 
suggested by Wills \etal (1993) in their study of AGNs.

Both the H$\alpha$ and \ion{He}{2} $\lambda 4686$ emission lines  
were, on average, asymmetric; they were slightly 
blue-shifted with respect to the central line wavelength (taking into account 
the systemic velocity). In the combined 
spectrum of all six nights, the H$\alpha$ line is blue-shifted by 
$55 \pm 9$ \kms, 
the H$\beta$ line by $61 \pm 12$ \kms, the narrow component of the 
\ion{He}{1} $\lambda 6678$ line by $82 \pm 22$ \kms,
and the \ion{He}{2} $\lambda 4686$ line by $41 \pm 13$ \kms. 
We note that a similar phenomenon is often observed in the emission lines 
from quasars and in some cataclysmic variables, and can be attributed 
to the expansion of clouds 
or to a gas outflow, when the receding gas is occulted from our view.

\subsubsection{Optically thin narrow-emission-line region}

Owing to the well-determined orbital parameters, the upper limit to 
the size of the accretion disk in GRO J1655$-$40 is reasonably well 
constrained.  We have found that the velocities inferred from the 
narrow widths of the H$\alpha$, H$\beta$ and 
\ion{He}{2} $\lambda 4686$ lines were significantly lower than 
the velocities expected from a Keplerian disk contained in the Roche lobe 
of the compact star.

One might argue that the low FWHM of the narrow emission 
lines was due to their formation in a  
circumbinary disk (Artymowicz \& Lubow 1996). If this were true, 
no significant velocity shifts from the center-of-mass 
systemic velocity should be seen over an orbital cycle. 
However, Soria \etal (1998) have measured the radial velocity shifts 
of the line wings at one-quarter of the maximum intensity above the continuum, 
and found a sinusoidal variation consistent with the one observed 
in 1996 June, which was interpreted 
as the modulation due to orbital motion of the compact star. 
Moreover, the profiles of these 
lines changed substantially from night to night. In the case of 
H$\alpha$, the line profile was single-peaked on four nights, but 
double-peaked on the other two, with a velocity separation larger than 
expected from a circumbinary disk. 
We can therefore rule out the possibility of emission from 
a circumbinary disk. 

We interpret the narrow lines, instead, as emission 
from a spheroidal or extended optically thin region high above 
the accretion disk (at lower rotational velocity). We notice 
that some cataclysmic variables show narrow, often single-peaked 
Balmer and \ion{He}{2} emission lines, yielding rotational velocities 
much slower than Keplerian. For example, there is strong evidence 
of an optically thin 
line-emitting region extended well above the disk plane in 
the eclipsing system PG 1012$-$029 (Honeycutt, Schlegel \& Kaitchuck 1986).
The larger FWHM of the high-ionization \ion{He}{2} 
line suggests that it was probably emitted closer to the disk plane 
than H$\alpha$, and/or at slightly smaller radii, where we expect the 
irradiation from the central source to be stronger. 

The night-to-night variability in the line profiles (Figure~\ref{n_stack94}) 
might reflect changes in the line-of-sight opacity of the optically thin 
region. Opacity variations can be caused by 
matter inhomogeneities (i.e.\ moving clouds) in the line emission 
region or variable obscuration along the light path.
Furthermore, the variability between single-peaked and 
double-peaked profiles indicates that the extended 
line-emitting region was not in steady state; for instance, 
the profiles would tend to become broader and more ``disk-like'' 
(as those observed in 1996 June) as the thin clouds dissipated, 
and less ``disk-like'' when the emission from the clouds was stronger.

The presence of optically thin line-emitting gas well above the disk 
plane may be explained in various ways: some accreting 
matter might not yet have settled down onto the disk, after the end of the 
strong hard X-ray flare observed by BATSE in 1994 August. We shall argue 
in \S6.4.2 that the outer disk is likely to be 
disrupted and puffed up into a thick torus during a hard X-ray flare. 
Some of the gas in the extended 
narrow-line region may also have been the result of bipolar outflows 
from the disk surface, as suggested by the systematic blue-shift 
and by the detection of strong winds in the broad-line region.

The extended, optically thin narrow-line-region model is valid only 
if the UV and soft X-ray flux is not too high. Soft irradiating flux 
of order of the Eddington limit at distances $R < 10^{12}$ cm would yield  
an ionization parameter $U > 10^4$ in the thin narrow-line region, 
for a gas with densities $n_{\scriptsize {\mbox {H}}} < 10^{13}$ cm$^{-3}$. 
This would imply temperatures $\simgt 10^7$ K, 
difficult to reconcile with the presence of lines such as H$\alpha$.
Although no UV and soft X-ray data were obtained at that epoch, the fact 
that we detected only low-ionization metal lines (for example from 
\ion{N}{2} and \ion{O}{2}) implies that the 
soft flux could not have been too strong in the 1994 X-ray outburst.

\subsection{Spectral transition on 1994 September 6} 

\subsubsection{General features}

\begin{figure} 
\begin{tabular}{c}
\epsfxsize=95mm\epsfbox{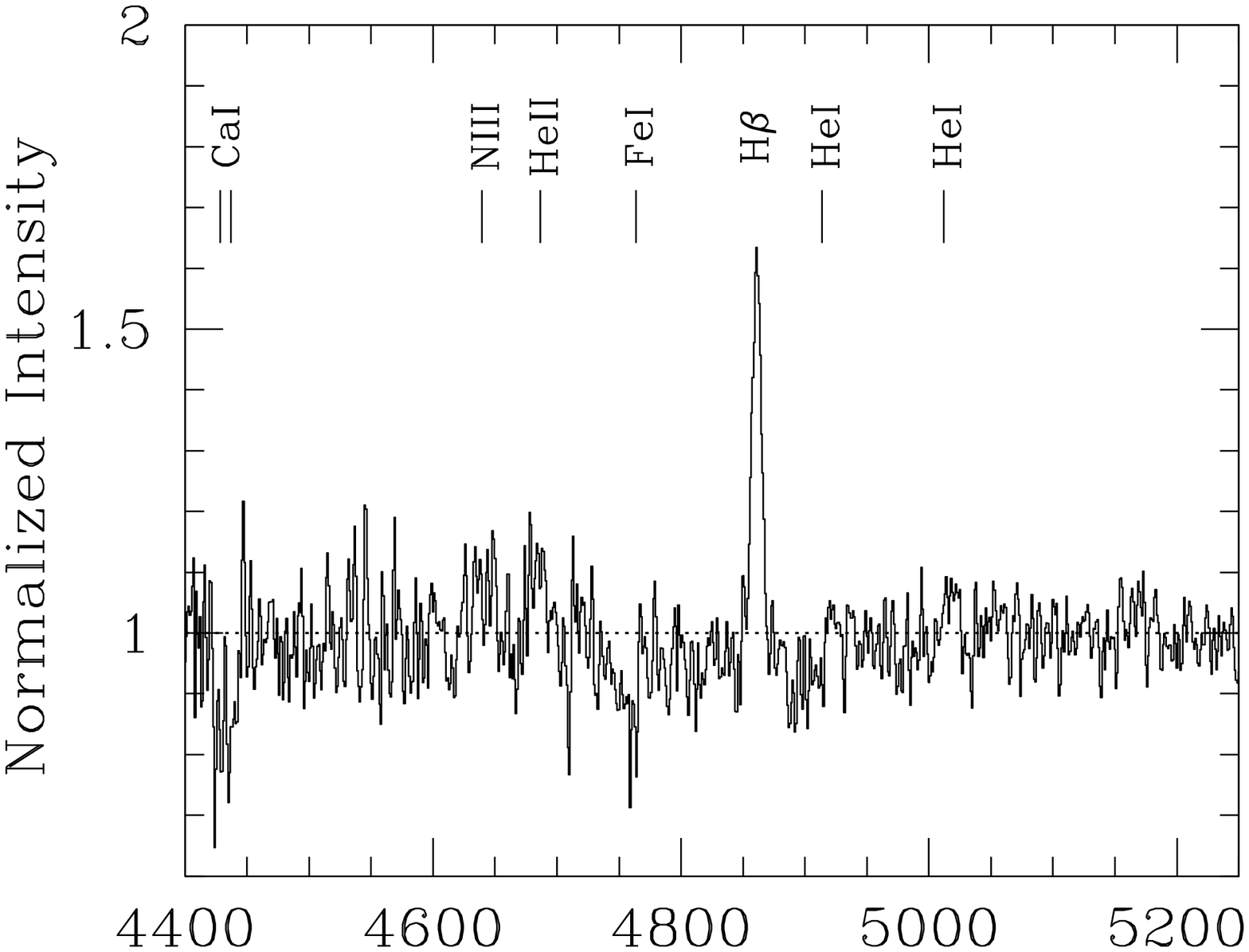}\\
\epsfxsize=95mm\epsfbox{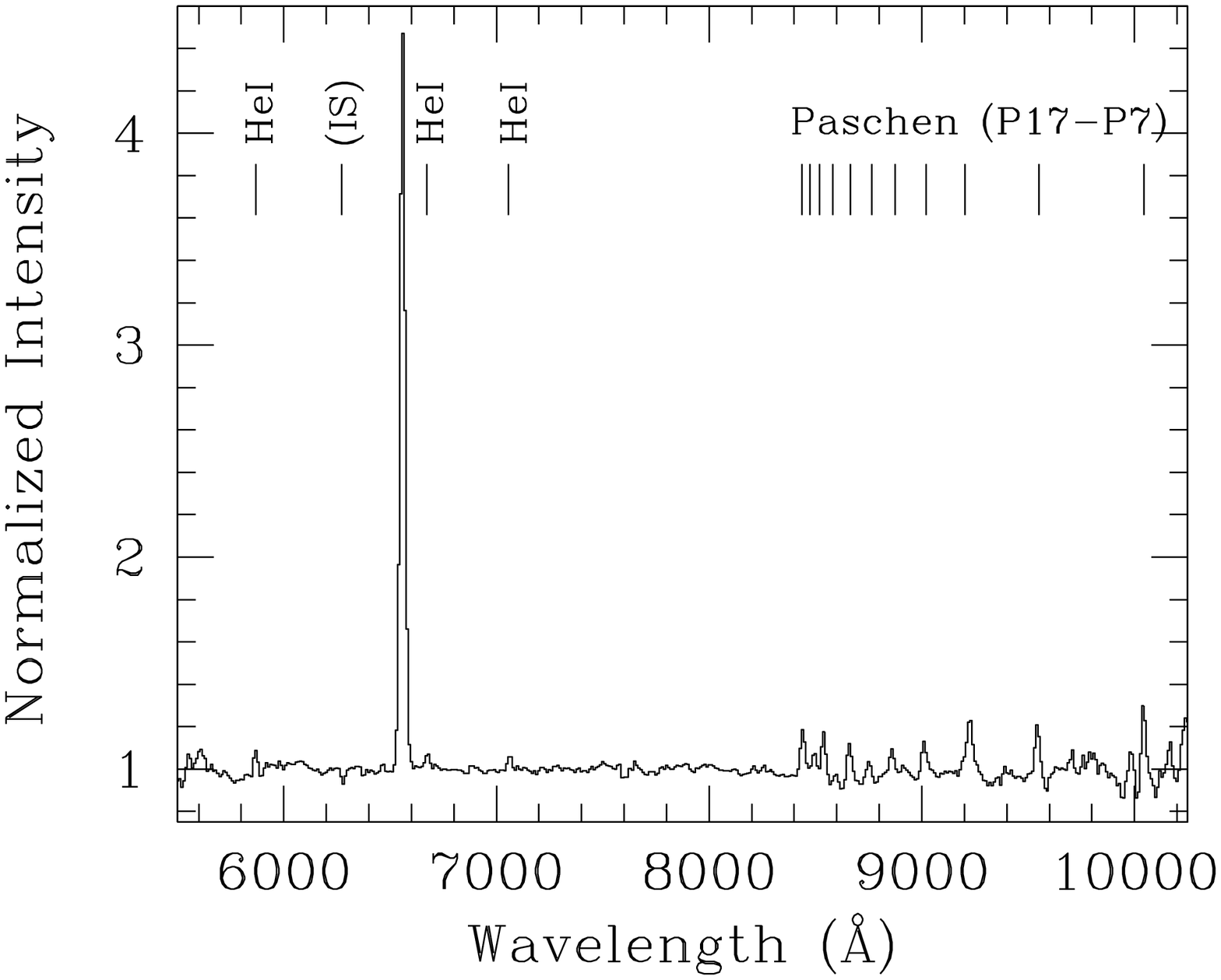}\\
\end{tabular} 
\caption{Top panel: medium-resolution blue spectrum obtained 
on 1994 September 6. Bottom panel: low-resolution red spectrum 
obtained at the same time.
Wavelengths are vacuum heliocentric and the intensity is 
normalized to the continuum. The flat-topped metal emission lines and the 
broad Balmer absorption troughs observed in the previous nights 
had disappeared.  High-ionization lines such as \protect\ion{He}{2} and Bowen 
\protect\ion{N}{3} had become much weaker; 
cf.\ Figure~\protect\ref{total94_hb1}. 
The narrow H$\beta$ emission line had become stronger, but the 
H$\alpha$$/$H$\beta$ EW ratio had also increased dramatically. 
The \protect\ion{H}{1} Paschen lines were seen in emission.
The dramatic change in the optical spectrum between September 4 and 6 
coincided with a surge in the $20-100$ keV X-ray flux measured by 
BATSE, and was followed by a major ejection of radio-emitting 
plasma on September 9.
}
\label{06sep}
\end{figure}

A sharp increase in the hard X-ray flux was detected by BATSE 
around September 5 (Harmon \etal 1995). The  
broad \ion{N}{2}, \ion{N}{3} and \ion{O}{2} emission lines 
disappeared from the blue spectrum obtained on September 6 
(Figure~\ref{06sep}), and 
\ion{He}{2} $\lambda 4686$ and the Bowen \ion{N}{3} lines were 
much weaker than in the previous nights. H$\alpha$ 
and H$\beta$ were instead much stronger: 
for H$\alpha$, EW $= 100 \pm 5$ \AA, and for H$\beta$, 
EW $=8.0 \pm 0.5$ \AA. The low-resolution of the red spectrum does not 
allow for a reliable determination of the FWHM of H$\alpha$; for H$\beta$, 
FWHM $=  530 \pm 50$ \kms. The H$\alpha$$/$H$\beta$ EW ratio increased 
by a factor of 4 between September 4 and September 6.
Other lines clearly seen in emission are 
the Paschen series of \ion{H}{1} and several 
\ion{He}{1} lines ($\lambda 4922$, $\lambda 5016$, 
$\lambda 5876$, $\lambda 6678$, $\lambda 7065$). All these emission 
lines were single-peaked; unfortunately, the low S/N of the blue spectrum 
and the low resolution of the red 
spectrum do not permit an accurate determination of their strengths 
and widths.  

\begin{figure}  
\epsfxsize=145mm\epsfbox{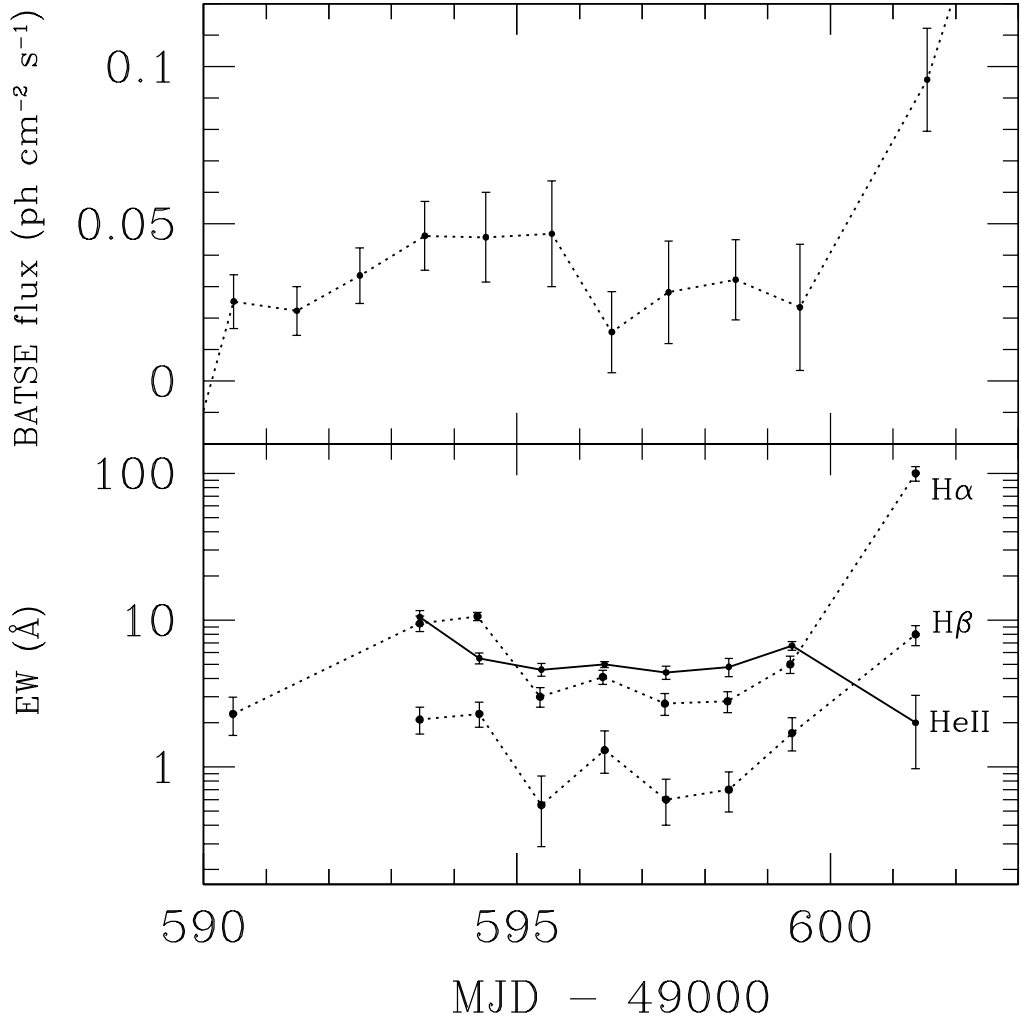} 
\caption{Top panel: hard ($20-100$ keV) X-ray flux, measured by BATSE 
during our 1994 August--September observations. Bottom panel: equivalent width 
of the H$\alpha$, H$\beta$ and \protect\ion{He}{2} 
$\lambda 4686$ emission lines during the same epoch. The EW of H$\alpha$ on  
MJD 49601.36 (September 6) was $100 \pm 5$ \AA, a factor of 20 stronger 
than on the previous nights. 
A similar correlation between H$\alpha$ emission and hard X-ray flux was 
observed in 1996 June.
The EW of H$\alpha$ on MJD 49590.47 has been determined from  a spectrum 
taken for us on August 26 by Gary Da Costa. The values of 
EW on MJD 49593.47 have been  
obtained from a spectrum taken for us on August 29 by Raffaella Morganti.}
\label{batse_EW94}
\end{figure}

The sudden increase in the Balmer emission between September 4 and 
September 6 coincided with the sharp increase in the hard X-ray flux.  
The EWs (in absolute value) of H$\alpha$, H$\beta$ 
and \ion{He}{2} $\lambda 4686$ are shown in Figure~\ref{batse_EW94} 
compared with the hard X-ray flux measured by BATSE. 
The Balmer EWs and the hard X-ray flux appear well correlated, 
as in the 1996 June observations. 
Such correlation is not seen for \ion{He}{2} $\lambda 4686$. 
In fact, its intensity dropped when the the hard X-ray flux rose.
 
Shrader \etal (1996) also noticed the dramatic increase in the 
H$\alpha$ emission and the disappearance of \ion{He}{2} $\lambda 4686$ 
around TJD 9600, after the hard X-ray flux surge 
(see their Fig.\ 1). Their low-resolution spectrum obtained on 
September 8 showed only emission from the low-ionization lines 
\ion{H}{1}, \ion{He}{1} and \ion{O}{1}.

Optical spectra taken by Bianchini \etal (1997) on 1994, August 12 -- 15, 
during the decline of an earlier hard X-ray flare, showed 
emission almost as strong as that in our September 6 spectrum from 
low-ionization lines like H$\alpha$ (mean EW $= 68.4$ \AA) and H$\beta$ 
(mean EW $= 7.4$ \AA). The H$\alpha$$/$H$\beta$ EW ratio was 9.2, 
a factor of 3 higher than on September 4, and slightly lower than on 
September 6. The high-ionization 
\ion{He}{2} $\lambda 4686$ (mean EW $= 5.3$ \AA) and 
\ion{N}{3} $\lambda \lambda 4641, 4642$ (mean EW $= 5.9$ \AA) lines 
were also present, showing disk or 
disk-wind signatures (see Fig.\ 2 in Bianchini \etal 1997). From 
the peak separation and the width of the flat-topped profiles, 
we can infer that the outer-disk radius was 
$\simlt (1.3 \pm 0.1) \times 10^{11}$ cm. Broad absorption troughs were 
not seen in those spectra. 

We suspect that the conditions on August 12 -- 15 were intermediate 
between those we observed before and after September 5. 
In the period of low X-ray activity between August 15 
and September 5, a geometrically thin accretion disk was probably re-forming 
and re-expanding to its tidal radius after having been disrupted in 
the previous hard X-ray flare, only to be then ``evaporated'' again 
into an extended atmosphere or cocoon by the following flare.

\subsubsection{Optically thick Balmer emission}

As the photoionization cross-section for 
hydrogenic ions falls approximately as $\nu^{-7/2}$ for photon 
energies larger than the threshold energy (13.6 eV for hydrogen) 
(see Osterbrock 1989), the hard ($20-100$ keV) X-ray photons 
contribute little to the ionization level of the gas, compared with the 
UV and soft X-ray photons. 
However, Compton heating by the hard X-rays can 
alter the disk structure and drive a disk-wind 
(e.g.\ Begelman, McKee \& Shields 1983; Idan \& Shaviv 1996). In the 
extreme conditions, the dense disk-wind would create a cocoon 
surrounding the accretion disk; in this situation one may consider 
the accretion disk being practically evaporated into a thick atmosphere. 
The absence of the characteristic broad absorption troughs at H$\alpha$ 
and H$\beta$ and of broad disk-like emission lines in our September 6 spectrum 
suggests the absence of an exposed, geometrically thin, optically thick 
accretion disk. 

An increase in the optical depth 
of the corona could have caused the increase in Balmer line emission and 
the disappearance of the high-ionization lines. 
The higher H$\alpha$$/$H$\beta$ EW ratio measured during the 1994 August and 
September hard X-ray flares, and the Paschen emission lines observed on 
September 6, indicate that the corona was optically 
thick in the Balmer series. When this situation occurs, every H$\beta$ 
photon is resonantly absorbed by a nearby hydrogen atom in the 
$n=2$ state; the process of emission and reabsorption is repeated until 
the H$\beta$ photon splits into H$\alpha$ + P$\alpha$ photons, which 
will both escape (``Case C'' recombination, see e.g.\ McCray 1996). 

\begin{figure}  
\epsfxsize=145mm\epsfbox{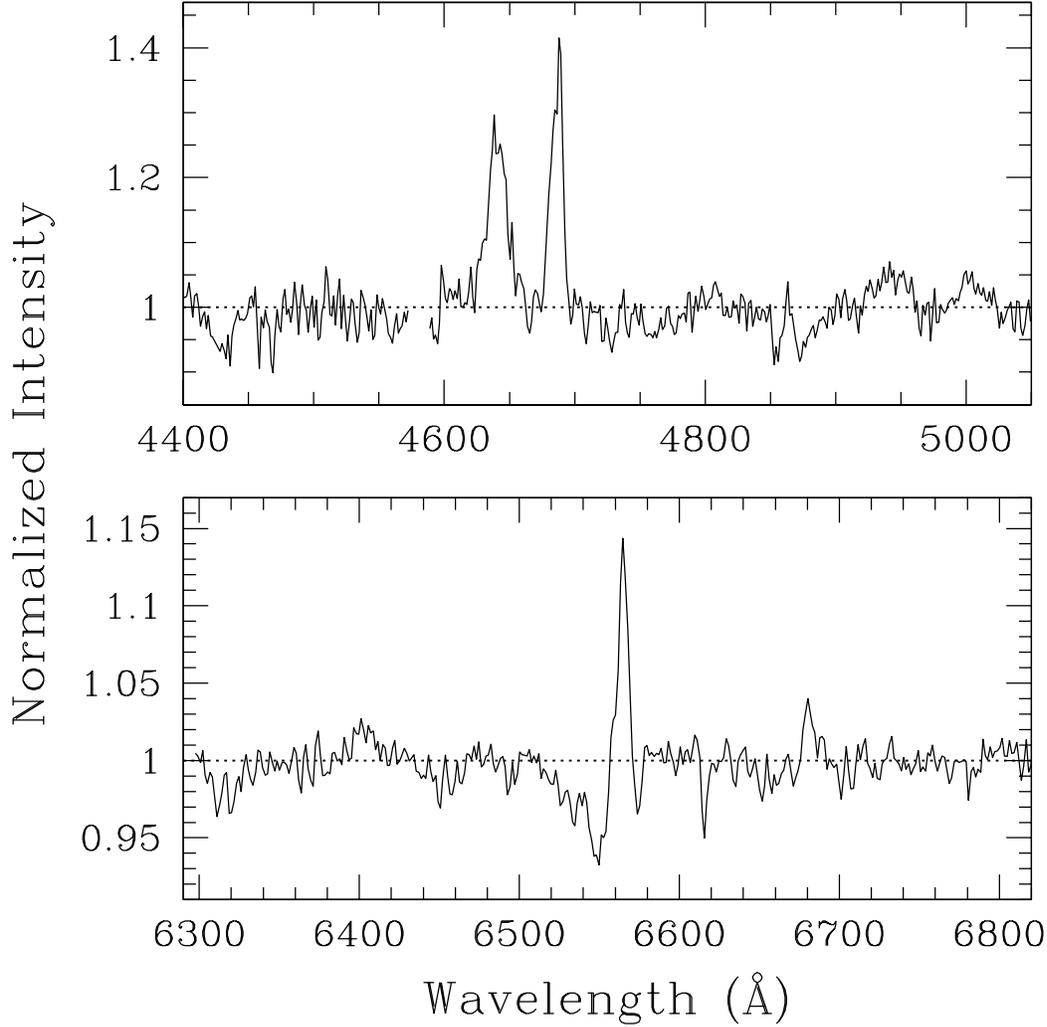}
\caption{Top panel: medium-resolution blue spectrum taken 
with the 3.9~m Anglo-Australian Telescope 
on 1994 September 27 (HJD 2449622.867 $=$ MJD 49622.367). Bottom panel: 
medium-resolution red spectrum from the same night 
(HJD 2449622.902 $=$ MJD 49622.402). 
These spectra, obtained from the AAT archive, show that two weeks after 
the end of the hard X-ray flare 
(cf.\ Figure 2 in Hjellming 1997) the system had returned to a state 
very similar to the one we observed before September 5. 
Cf.\ Figures~\protect\ref{total94_ha} and \protect\ref{total94_hb1} 
for an identification of the main features.}
\label{Greenhill94}
\end{figure}

After the end of the 1994 August hard X-ray 
flare, the extended cocoon probably became optically thin to the Balmer 
lines (``Case B'' recombination), as observed on August 30 -- September 4. 
The disk was ``puffed up'' again by the hard X-ray flare 
after September 4, so that Balmer lines were emitted via Case C 
recombination. Archival spectra obtained from the Anglo-Australian 
Observatory, taken on September 27 (Figure~\ref{Greenhill94}), show 
features almost identical to those observed on August 30 -- September 4 
(cf.\ Figures~\ref{total94_ha} and \ref{total94_hb1}), suggesting that 
the cocoon once again became optically thin after the end of the 
second hard X-ray flare.

The simultaneous decrease in the observed strength of the 
\ion{He}{2} and Bowen \ion{N}{3} lines at the time of the hard X-ray 
increase can perhaps be attributed to a 
geometric effects. As GRO~J1655$-$40 has a high orbital 
inclination, a thick outer-disk rim, inflated by hard X-ray irradiation, 
may occult the inner region where the high-ionization lines are formed, 
and explain their weakness in the September 6 spectrum.   

Finally, we note that a major episode of plasma ejection occurred on 
1994 September 9, inferred from the radio observations 
(Harmon \etal 1995; see also Wu \& Hunstead 1997), a few days after 
the change in the optical and hard X-ray spectra. It was suggested 
(Fabian \& Rees 1995) that there may be a connection between the 
production of collimated jets and the presence of an extended hot, 
high pressure atmosphere and a quasi-spherical accretion flow in 
the centers of elliptical galaxies, the usual hosts for radio-loud 
emission. Whether the same mechanism was a catalyst 
for the collimated ejections observed in the radio-loud outburst 
of GRO J1655$-$40 in 1994 August -- September is an issue worth 
investigating.

\section{SUMMARY}

We have obtained optical spectra of the soft X-ray transient 
GRO~J1655$-$40 during different energy states (quiescence, 
high-soft and hard outburst) between 1994 August 
and 1997 June. 

Spectra taken one day after the observed brightening 
in $V$ in 1996 April did not at that stage show evidence of emission 
lines from an accretion disk. The main spectral features were narrow 
absorption lines from the secondary star. The optical brightening 
was probably due to an increase in the continuum emission from 
the disk; emission lines could not be formed at that stage, 
in the absence of X-ray irradiation.

For the 1996 -- 97 high-soft state, we have identified characteristic 
features such as broad absorption lines at H$\alpha$ and H$\beta$, 
partly filled by double-peaked emission lines. We argue that the broad 
absorption was formed in the hot inner disk, and the double-peaked 
\ion{He}{2} $\lambda 4686$ emission originated in a temperature-inversion 
layer on the disk surface, created by the soft X-ray irradiation.
The observed rotational velocities suggest that the disk was probably 
extended beyond its tidal radius.
We have also found that the double-peaked H$\alpha$ emission was associated 
with the hard X-ray flux, suggesting that it was probably emitted at 
comparable radii but from 
deeper layers (at higher optical depth) than \ion{He}{2} $\lambda 4686$.
We note that the Balmer lines were not emitted uniformly from the whole 
disk surface, but appeared to come only from a double-armed 
region, possibly the effect of tidal density waves or shocks on the disk.

We have identified three classes of lines in the spectra taken  
in 1994 August -- September before the onset of a hard X-ray flare: 
broad absorption, broad (flat-topped) emission and narrow emission.
The narrow emission line profiles cannot be explained by a conventional 
thin accretion disk model. We speculate that the system was in a transient 
state in which the accretion disk had an extended, optically 
thin cocoon and significant matter outflow, which would also help to 
explain the systematic blue-shift of the narrow emission lines and the 
flat-topped profiles of the broad emission lines.

After the onset of a hard X-ray flare, the disk signatures disappeared, 
and strong H$\alpha$ and Paschen emission was detected, suggesting 
that the cocoon became optically thick to the Balmer lines. 
High-ionization lines disappeared or weakened.
Two weeks after the end of the flare, the cocoon appeared to be 
once again optically thin.

We thank: David Buckley, Gary Da Costa, Paul Francis, Charlene Heisler 
and Raffaella Morganti for 
taking some of the spectra; Alan Harmon and Shuang-Nan Zhang 
for the BATSE data and discussions; and Geoff Bicknell, 
Ralph Sutherland, Martijn de Kool, Stefan Wagner and John Greenhill 
for comments and discussions. R.~W.~H. acknowledges financial assistance from 
the Australian Research Council, and thanks his co-observers in 1994 
August -- September, Max Pettini, Dave King and Linda Smith, for their 
tolerance in allowing the ToO observations to displace some of the 
scheduled observing program. K.~W. acknowledges support from 
the ARC through an Australian Research Fellowship and the support 
from S.~N.~Zhang for the visits to NASA-MSFC and UAH.

\clearpage

\begin{deluxetable}{lrc}
\footnotesize
\tablecaption{Physical parameters for GRO J1655$-$40}
\tablewidth{0pt}
\tablehead{
\colhead{} & \colhead{} & \colhead{Ref.}
}
\startdata
Distance  & $d = 3.2 \pm 0.2$ kpc & a \nl
 & $3 < d < 5$ kpc & b \nl
Binary period & $P = 2.62157 \pm 0.00015$ d  & c \nl
 & $P = $2.62168(14) d & d \nl
Inclination angle & $i = 69^{\circ}.5 \pm 0^{\circ}.08$  & c \nl
 & $63^{\circ}.7 < i < 70^{\circ}.7$ & d \nl
Mass ratio & $q = 0.334 \pm 0.009$  & c \nl
 & $0.337 < q < 0.436$ & e \nl
Mass function & $f_{\rm X} = 3.24 \pm 0.09$ $M_{\odot}$  & c \nl
 & $f_{\rm X} = 2.73 \pm 0.09$ $M_{\odot}$  & e \nl
 & $1.93 < f_{\rm X} < 2.67$ $M_{\odot}$ & f \nl
Mass of primary & $M_{1} = 7.02 \pm 0.22$ $M_{\odot}$  & c \nl
 & $5.5 < M_{1} < 7.9$ $M_{\odot}$ & e \nl
 & $4.1 < M_{1} < 6.6$ $M_{\odot}$ & f \nl
Mass of secondary & $M_{2} = 2.34 \pm 0.12$ $M_{\odot}$  & c \nl
 & $1.7 < M_{2} < 3.3$ $M_{\odot}$ & e \nl
 & $1.4 < M_{2} < 2.2$ $M_{\odot}$ & f \nl
Systemic velocity & $\gamma = -142.4 \pm 1.6$ \kms  & c \nl
 & $\gamma = -141.9 \pm 1.3$ \kms  & e \nl
 & $-153 < \gamma < -143$ \kms & f \nl
Quiescent $V$ mag & $17.3$  & g \nl

\tablenotetext{a}{Hjellming \& Rupen 1995}
\tablenotetext{b}{Tingay \etal 1995}
\tablenotetext{c}{Orosz \& Bailyn 1997}
\tablenotetext{d}{van der Hooft \etal 1998}
\tablenotetext{e}{Shahbaz \etal 1999}
\tablenotetext{f}{Phillips \etal 1999}
\tablenotetext{g}{Bailyn \etal 1995}

\enddata
\end{deluxetable}

\clearpage

\begin{deluxetable}{lccccc}
\footnotesize
\tablecaption{Our spectroscopic observations of GRO~J1655$-$40}
\tablewidth{0pt}
\tablehead{
\colhead{Date} & \colhead{Epoch}  
& \colhead{Phase range} & \colhead{Wavelength range} 
& \colhead{Resolution} \nl
& \colhead{(HJD - 2449000)}  & & \colhead{(\AA)} 
& \colhead{(\AA\ FWHM)}
}
\startdata
\multicolumn{5}{c}{AAT 3.9~m} \nl
\hline
1994 August 26 & 590.969--590.976 & 0.359--0.362 & 5645--7240 & 3.5 \nl
1994 August 29 & 593.947--593.954 & 0.495--0.498 & 3780--7450 & 7.3 \nl
1994 August 30 & 594.866--594.883 & 0.846--0.852 & 6280--6825 & 1.3 \nl
& 594.897--594.909 & 0.858--0.862 & 4432--5051 & 1.3 \nl
1994 August 31 & 595.863--595.877 & 0.226--0.231 & 6280--6825 & 1.3 \nl
& 595.885--595.899 & 0.234--0.240 & 4432--5051 & 1.3 \nl
1994 September 1 & 596.867--596.883 & 0.609--0.615 & 6280--6825 & 1.3 \nl
& 596.888--596.911 & 0.617--0.626 & 4432--5051 & 1.3 \nl
1994 September 2 & 597.857--597.870 & 0.987--0.992 & 6280--6825 & 1.3 \nl
& 597.872--597.885 & 0.992--0.997 & 4432--5051 & 1.3 \nl
1994 September 3 & 598.860--598.873 & 0.369--0.374 & 6280--6825 & 1.3 \nl
& 598.875--598.894 & 0.375--0.382 & 4432--5051 & 1.3 \nl
1994 September 4 & 599.860--599.873 & 0.751--0.756 & 6280--6825 & 1.3 \nl
& 599.882--599.895 & 0.759--0.764 & 4432--5051 & 1.3 \nl
1994 September 6 & 601.861--601.864 & 0.514--0.515 & 5500--11000 & 20 \nl
&&& 3925--5500 & 2.5 \nl
\hline
\multicolumn{5}{c}{ANU 2.3~m} \nl
\hline
1996 April 20 & 1194.259--1194.303 & 0.485--0.501 & 3600--9300 & 4.8 \nl
1996 April 21 & 1195.096--1195.151 & 0.804--0.824 & 3600--9300 & 4.8 \nl
1996 June 8 & 1242.980--1243.327 & 0.069--0.202 & 4360--5320 & 1.3 \nl
	     & 1242.908--1243.327 & 0.042--0.202 & 6300--7250 & 1.3 \nl
1996 June 9 & 1243.893--1244.285 & 0.418--0.567 & 4150--5115 & 1.3 \nl
&&&6300--7250 & 1.3 \nl
1996 June 10 & 1244.868--1245.292 & 0.790--0.951 & 4150--5115 & 1.3 \nl
&&&6300--7250 & 1.3 \nl
1996 June 11 & 1245.885--1246.153 & 0.178--0.280 & 4150--5115 & 1.3 \nl
&&&6300--7250 & 1.3 \nl
1996 June 12 & 1246.877--1247.293 & 0.556--0.715 & 4150--5115 & 1.3 \nl
&&&6300--7250 & 1.3 \nl
1996 June 17 & 1252.029--1252.052 & 0.521--0.530 & 6300--7250 & 1.3 \nl
1997 June 14 & 1614.010--1614.092 & 0.599--0.630 & 4150--5115 & 1.3 \nl
&&&6300--7250 & 1.3 \nl
1997 June 15 & 1615.056--1615.243 & 0.998--0.069 & 4150--5115 & 1.3 \nl
&&&6300--7250 & 1.3 \nl
\enddata
\end{deluxetable}

\clearpage

\begin{deluxetable}{lcc}
\footnotesize
\tablecaption{Main permitted lines detected during the 1994 outburst}
\tablewidth{0pt}
\tablehead{
\colhead{Line} & \colhead{EW (\AA)} & \colhead{FWHM (\kms)} 
}
\startdata
\multicolumn{3}{c}{BROAD ABSORPTION LINES} \nl
\hline
H$\beta$ $\lambda 4861$ & $-4 \pm 1$ & $3000 \pm 600$ \nl
H$\alpha$ $\lambda 6562$ & $-6 \pm 1$ & $2700 \pm 500$ \nl
\hline 
\multicolumn{3}{c}{BROAD EMISSION LINES} \nl
\hline
\ion{O}{2} $\lambda 4452$ & $0.7 \pm 0.2$ & $1350 \pm 150$ \nl
\ion{N}{3} $\lambda 4515$ & $0.7 \pm 0.2$ & $950 \pm 50$ \nl
\ion{He}{2} $\lambda 4542$ & $0.5 \pm 0.2$ & $800 \pm 50$ \nl
\ion{O}{2} $\lambda \lambda \lambda 4602,4609,4613\tablenotemark{a}$ 
	& $2.1 \pm 0.2$ & $1400 \pm 150$ \nl
\ion{N}{3} $\lambda \lambda 4641,4642\tablenotemark{b}$ 
& $5.0 \pm 0.5$ & $1050 \pm 100$ \nl
\ion{O}{2} $\lambda \lambda 4941, 4943
\tablenotemark{c}$ & $1.9 \pm 0.1$ & $1750 \pm 50$ \nl
\ion{N}{2} $\lambda 5005$ & $1.0 \pm 0.1$ & $1400 \pm 50$ \nl
\ion{He}{1} $\lambda 6678\tablenotemark{d,e}$ 
& $0.9 \pm 0.1$ & $950 \pm 100$ \nl
\hline 
\multicolumn{3}{c}{NARROW EMISSION LINES} \nl
\hline
H$\alpha$ $\lambda 6562$ & $5.3 \pm 0.2$ & $450 \pm 20$ \nl
H$\beta$ $\lambda 4861$ & $1.1 \pm 0.1$ & $550 \pm 30$ \nl
\ion{He}{2} $\lambda 4686$ & $5.2 \pm 0.2$ & $540 \pm 20$ \nl
\ion{He}{1} $\lambda 6678\tablenotemark{e,f}$ 
& $0.9 \pm 0.1$ & $380 \pm 40$ \nl
\tablenotetext{a}{Possible contribution also from 
\ion{N}{2} $\lambda \lambda \lambda 4601,4607,4614$ and (less likely) 
from \ion{N}{4} $\lambda 4606$.}
\tablenotetext{b}{after deblending a small contribution from 
\ion{N}{2} $\lambda 4631$; possible contribution also from 
\ion{N}{3} $\lambda 4634$.}
\tablenotetext{c}{after deblending a small contribution from 
\ion{He}{1} $\lambda 4922$.}
\tablenotetext{d}{``broad'' component of the line.}
\tablenotetext{e}{Possible contribution also from \ion{He}{2} $\lambda 6683$.}
\tablenotetext{f}{``narrow'' component of the line.}

\enddata
\end{deluxetable}

\clearpage

\clearpage

\clearpage

\clearpage

\clearpage

\clearpage

\clearpage

\clearpage

\clearpage

\clearpage

\clearpage

\clearpage

\clearpage

\clearpage

\clearpage

\clearpage

\clearpage

\clearpage

\clearpage

\end{document}